\documentclass[apj]{emulateapj}  

\usepackage{graphicx} 
\usepackage{apjfonts}

\newcommand{\ia}{\'{\i}}

\shorttitle{DOWNFLOWS IN SUNSPOT UMBRAL DOTS}
\shortauthors{ORTIZ ET AL.}

\begin{document}

\title{Downflows in sunspot umbral dots}
\author{A. Ortiz$^1$}
\author{L.R. Bellot Rubio$^2$}
\author{L. Rouppe van der Voort$^1$}
\affil{$^1$ Institute of Theoretical Astrophysics, University of Oslo, P.O. Box 1029 
Blindern, N-0315 Oslo, Norway; ada@astro.uio.no, }
\affil{$^2$ Instituto de Astrof\ia sica de Andaluc\ia a (CSIC), Apdo.\ 3004, 18080 Granada, Spain}

\journalinfo{The Astrophysical Journal, 713:1282-1291, 2010 {\rm April} 20}
\submitted{Received 2010 February 13; accepted 2010 March 9; published 2010 April 1}

\begin{abstract}

We study the velocity field of umbral dots (UDs) at a resolution of
0\farcs14. Our analysis is based on full Stokes measurements of a pore taken with the Crisp Imaging Spectro-Polarimeter at the Swedish
1 m Solar Telescope. We determine the flow velocity at different
heights in the photosphere from a bisector analysis of the \ion{Fe}{1}
630~nm lines. In addition, we use the observed Stokes $Q$, $U$, and $V$
profiles to characterize the magnetic properties of these
structures. We find that most UDs are associated with strong
upflows in deep photospheric layers. Some of them also show
concentrated patches of downflows at their edges, with sizes of about
0\farcs25, velocities of up to 1000~m~s$^{-1}$, and enhanced net
circular polarization signals. The downflows evolve rapidly and have
lifetimes of only a few minutes. These results appear to validate
numerical models of magnetoconvection in the presence of strong
magnetic fields.

\end{abstract}


\section{Introduction}

At high angular resolution, sunspot umbrae exhibit small bright
features embedded in a darker, smoothly varying background. These
features are called umbral dots (UDs) and occur in essentially all
sunspots but also in pores \citep[for a review,
see][]{sobotka97}. Frequently a distinction is made between central
and peripheral UDs based on their location within the umbra
\citep[e.g.][]{gd86,rieth08a,sob09}. Central UDs do not show measurable
vertical flows and have similar magnetic field inclinations as the
surrounding umbra. In contrast, upflows and more horizontal magnetic
fields are detected in peripheral UDs.

The importance of UDs lies in the fact that they may be the signature
of convection in sunspots. We know that convective motions must exist
because radiation alone cannot explain the brightness of the umbra
\citep{deinzer65}. The question is whether these motions are field free
or magnetized. Recent simulations of magnetoconvection in a strong
background field \citep{simu06} suggest that the convective energy
transport inside the umbra is achieved in the form of narrow plumes of
rising hot plasma and strongly reduced magnetic fields. The plumes are
associated with bright structures that share many similarities with
real UDs, including their sizes (about 300 km) and lifetimes (30
minutes). The simulated UDs tend to be elliptical in shape, and many have
a central dark lane.

According to \cite{simu06}, the pile-up of material at the top of the
plumes increases the gas density and moves the $\tau=1$ level toward
higher layers, where the temperature is lower. This produces the dark
lanes of the simulated UDs. The upflows attain maximum velocities of
3-4 km~s$^{-1}$ just below the solar surface. Near $\tau=1$ they turn
horizontal and move in the direction of the dark lane, until the gas
returns to deeper layers along narrow channels. The downflows occur
mainly at the endpoints of the dark lanes, but also in the immediate
surroundings of the UDs.  They usually appear as tiny patches of
concentrated flows with velocities of up to 1200 m\,s$^{-1}$ at $z =
0$ km. The velocities at $\tau = 1$ are smaller because the large
opacity of the gas hides the layers where the flows are stronger
\citep{bhar10}.

The simulations of \cite{simu06} suggest that UDs are a natural
consequence of convection in a strong magnetic field, which seems to 
favor the monolithic sunspot model over the cluster model. However,
the basic physical process at work in the simulations, namely
overturning convection, has not yet been confirmed
observationally. Upflows are known to exist in UDs; what is missing is
a clear, unambiguous detection of return downflows.

The velocity field of UDs has been studied by a number of
authors. Most of them report negligible velocities in central UDs
\citep[e.g.,][]{schmidt94, rim08, rieth08a,sob09} and 
upflows ranging from 100 m s$^{-1}$ to 1000 m\,s$^{-1}$ in peripheral UDs
\citep[e.g.,][]{hector04,rim04,rieth08a, sob09}.
\citet{bar07} claim to have measured upflows of 400 m\,s$^{-1}$ 
and downflows of 300 m\,s$^{-1}$ in UDs, using the Universal
Birrefringent Filter at the Dunn Solar Telescope. This is a remarkable
result which might have been compromised by the fact that no adaptive
optics, phase diversity, or any postprocessing technique such as
speckle reconstruction were available at the time of the
observations. In any case, the downflows described by \citet{bar07}
seem to be different from those reported by \citet{simu06} in that
they extend over much larger areas. The same happens to the downflow
patches detected by \cite{bar09} using {\it Hinode} observations. 
Interestingly, the spectropolarimetric measurements of
\cite{rieth08a} and \cite{sob09}, also from {\it Hinode}, do not show
downward motions despite their better angular resolution (0\farcs3 versus
0\farcs6).

To detect the downflows predicted by the simulations, one would like
to have extremely high spatial resolution --higher than typically
achieved by current instruments-- and line spectra to probe the
velocity field at different heights in the atmosphere. Until now it
has been difficult to fulfill the two requirements simultaneously.
Here, we present the first spectropolarimetric observations of UDs
approaching a resolution of 0\farcs1. This unique data set, acquired at
the Swedish 1 m Solar Telescope (SST), allows us to investigate the flow
field and magnetic properties of UDs with unprecedented detail. The
measurements show localized patches of downflows at the edges of UDs,
providing firm evidence for magnetoconvection in sunspot umbrae.

The paper is organized as follows. The observations and the data
reduction are described in Section~\ref{data}. Section \ref{results}
deals with the morphology, flow field, temporal evolution, and
magnetic properties of UDs as derived from the observed Stokes
spectra. Finally, in Section~\ref{disc} we discuss our results and
compare them with previous works.

\section{Observations and data processing}
\label{data}
 
The observations were obtained with the CRisp Imaging
Spectro-Polarimeter (CRISP) at the SST
\citep{schar03} on La Palma (Spain). CRISP is based on a dual
Fabry P{\'e}rot interferometer (FPI) system similar to that described
by \citet{schar06}. The instrument has been designed to allow
diffraction-limited observations in the visible.

CRISP is equipped with three high-speed Sarnoff CCD cameras operating at
35~frames per second. Two of them are positioned behind the FPI, and
the third is located before the FPI but after the CRISP
pre-filter. This camera is used as an anchor channel for image
processing and is referred to as the wide-band channel. All the
cameras take exposures of 17 ms and are synchronized by means of an optical chopper. The image scale is 0\farcs071 pixel$^{-1}$, and the field
of view (FOV) is $\sim$71\arcsec$\times$71\arcsec. The CRISP
transmission profile has an FWHM of 6.4~pm at 630~nm.

The polarization modulation is performed by two nematic liquid crystal
(LC) variable retarders cycling through four different states. The time
needed to change between LC states is smaller than the CCD read-out
time (10~ms) by use of the transient nematic effect. A polarizing
beam-splitter in front of the narrow-band cameras divides the light
into two orthogonally polarized beams, which results in a significant
reduction of seeing-induced cross talk \citep{lites87}.

The data analyzed here were recorded on 2008 June 12 and correspond to
a pore located at S09 E30 ($\theta=31^\circ$, $\mu=0.85$). This pore 
is the naked umbra of a decaying sunspot (AR 10998) that showed a
rudimentary penumbra in previous days. CRISP was used to measure the
four Stokes profiles of the \ion{Fe}{1} 630.15 and 630.25 nm lines,
each sampled at 15 wavelength positions in steps of 4.8 pm, from
$-33.6$ to $+33.6$ pm. In addition, a continuum wavelength point was
observed. CRISP is capable of very fast wavelength tuning
($\lesssim$50~ms) within the same spectral region. We recorded 7
exposures per LC state, resulting in 28 exposures per wavelength
position. The total time for a complete wavelength scan of the two lines
was 26~s.

\begin{figure}
\begin{center}
\resizebox{6.5cm}{!}{\includegraphics[bb=70 0 526 410]{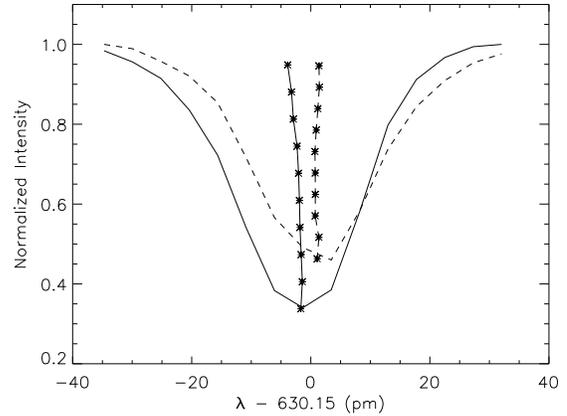}} 
\caption{\ion{Fe}{1} 630.15~nm intensity profiles and line bisectors 
of UDs showing upflows (solid) and downflows (dotted). The
corresponding spatial points are marked with crosses in
Figure~\ref{fig1}. The profiles have been normalized to their local
continua. The bisectors are indicated by asterisks and span the range of
intensity levels from 0\% to 90\% . The 80\% bisectors give a velocity
of $-1600$ m\,s$^{-1}$ for the upflow and $+810$ m\,s$^{-1}$ for the
downflow.
\label{bisector}}
\end{center}
\end{figure}

We followed the pore for 23 minutes during which the atmospheric
conditions were variable, with moments of excellent seeing. To improve
the spatial resolution, we used the adaptive optics system of the SST
\citep{schar03b} and the Multi-Object Multi-Frame Blind Deconvolution
image restoration technique
\citep[MOMFBD;][]{noort05}. All frames in a complete wavelength scan, 
amounting to 868 images per camera, or 2604 images in total (apart
from occasionally dropped frames due to system hick-ups), were
included in a MOMFBD restoration. The images were divided into
overlapping 64$\times$64 pixel subfields
(4$\farcs$5$\times$4$\farcs$5) and all images from each subfield were
processed as a single MOMFBD set.  We refer the reader to
\citet{noort08} for more details on the MOMFBD processing strategies
on similar multi-wavelength polarimetric scans.

The restored images were demodulated (pixel by pixel) and corrected
for telescope polarization using the telescope model developed by
\citet{sel05}. Details on the CRISP polarimetric calibration and 
the telescope model can be found in \citet{noort08}. After
demodulation, the two sets of orthogonal Stokes images recorded by the
narrow-band cameras were merged. Precise alignment between these
images is ensured by the pinhole calibration performed as part of the
MOMFBD processing. The final Stokes profiles were corrected for
residual $I$ to $Q$, $U$, and $V$ crosstalk by forcing the polarization
to be zero in the continuum. Another correction was applied to remove
the intensity gradient introduced by the CRISP pre filter (FWHM
0.44~nm).

The noise level in Stokes $Q$, $U$, and $V$ amounts to $2.3 \times
10^{-3}$, $1.6 \times 10^{-3}$, and $1.7 \times 10^{-3}$ of the
continuum intensity, respectively. This is roughly a factor of 2
higher than the noise level of \citet{schar08}, who used 4 times more
exposures. We confirmed that the spatial resolution achieved in our
Stokes maps is close to the diffraction limit of the SST by analyzing
intensity cuts of the smallest features in the images. We conclude
that the spatial resolution is basically limited by the pixel size,
0\farcs071, which slightly undersamples the theoretical diffraction
limit of the telescope ($\lambda/D=0\farcs13$ at 630~nm).

\subsection{Line parameters}

\begin{figure*}
\centering
\resizebox{.45\hsize}{!}{\includegraphics[bb= -50 -75 420 480]{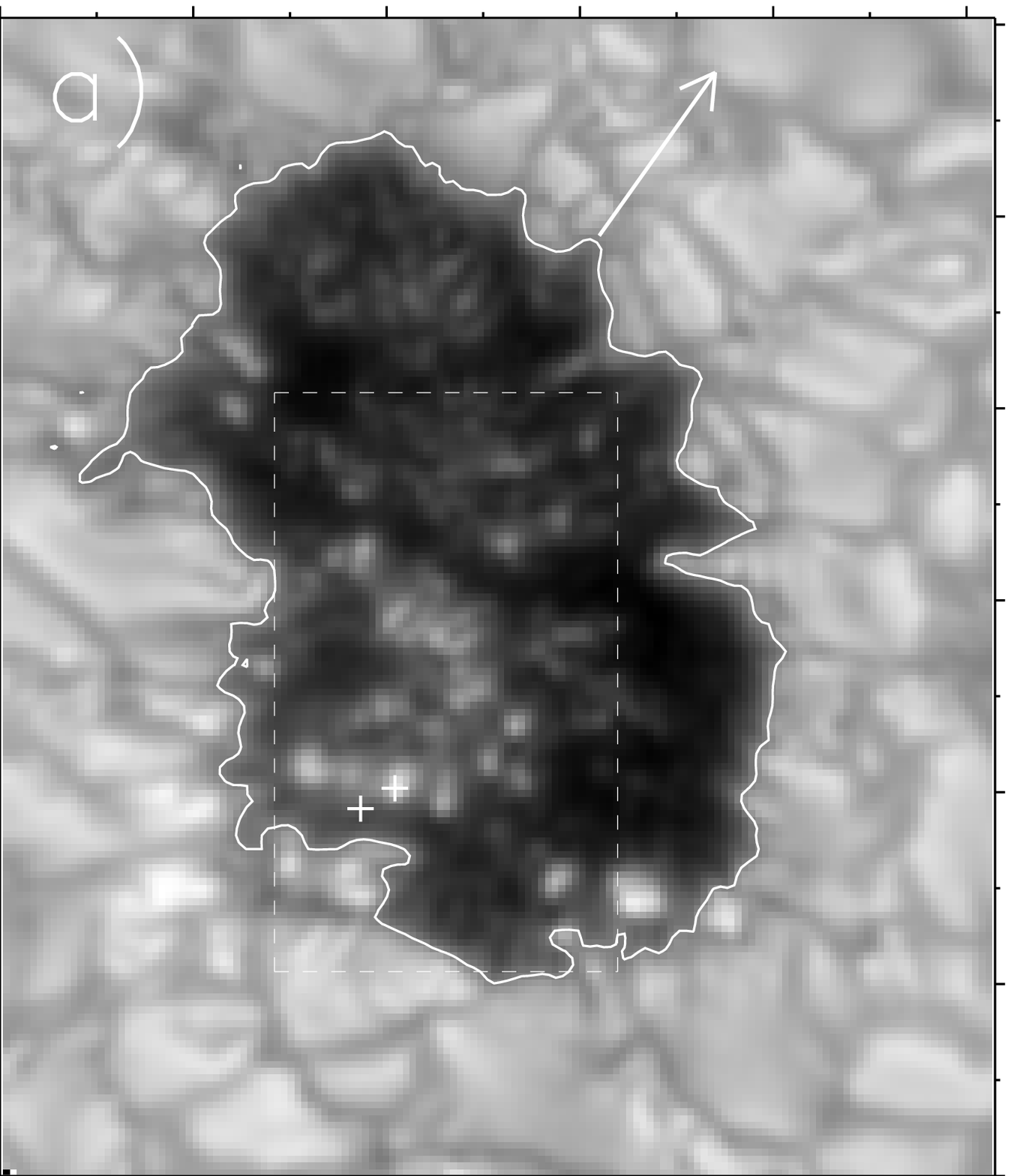}} 
\resizebox{.45\hsize}{!}{\includegraphics[bb= -50 -75 420 480]{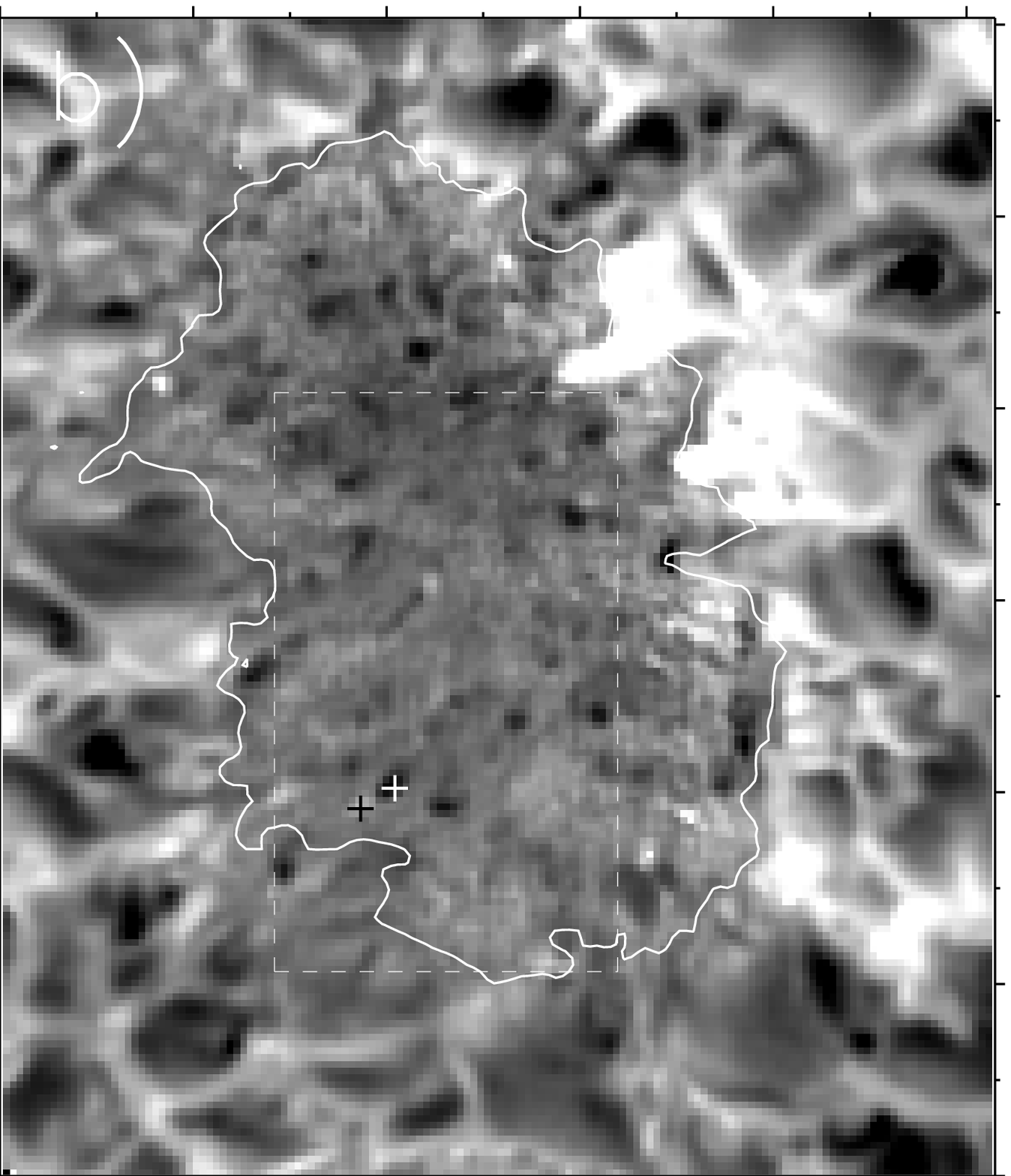}}
\caption{AR 10998 as observed by CRISP on 2008 June 12 at 10:33 UT. 
(a) Continuum intensity image from the narrow band channel of
CRISP. The contours outline the edge of the pore and the arrow marks
the direction to disk center. The rectangle encloses the region studied
in Figure~\ref{fig2}.  Note the complex network of magnetoconvective
structures inside the pore. There are UDs with and without dark
lanes. (b) LOS velocity map computed from the bisector position at the
$80\%$ intensity level. The gray scale ranges from $-1.5$ km\,s$^{-1}$
(black) to 1.5 km\,s$^{-1}$ (white).  Negative velocities indicate
blueshifts.
\label{fig1}}
\vspace*{-.1cm}
\end{figure*}

In this work, we use line bisectors to obtain the velocity field of
UDs. The bisector is the locus connecting the midpoints of horizontal
segments between the two wings of the intensity profile (see
Figure~\ref{bisector} for an example). We determined the bisector
positions at different intensity levels, from 0$\%$ (line core) to
80$\%$ (line wing near the continuum), with the help of the {\tt
bisec\_abs\_lin.pro} routine included in the Kiepenheuer-Institut
f\"ur Sonnenphysik IDL library. The line core position was computed by
means of a parabolic fit around the intensity minimum. The
calculations were performed only for \ion{Fe}{1} 630.15~nm, since the
red wing of \ion{Fe}{1} 630.25 nm is blended with a telluric O$_{2}$
line. The bisector positions were converted into Doppler velocities
assuming that the pore is at rest. More specifically, we averaged the
bisectors between the 40\% and 70\% intensity levels within the pore
and took this as the zero point of the velocity scale. The rms
fluctuation of the velocity in the pore is 320 m~s$^{-1}$, as computed
from the 80\% bisector. This is larger than the values reported by
\cite{franz09}, which we attribute to the higher spatial resolution of
our observations and the fact that we see clear flow structures
embedded in the dark umbral background. The systematic error of such a
velocity calibration is not larger than about 150 m~s$^{-1}$
\citep{reza06}.

The availability of spectrally resolved line profiles makes it
possible to probe the velocity field at different heights in the
photosphere. Roughly speaking, higher intensity levels correspond to
deeper layers. The continuum is formed at around $\log \tau = 0$,
whereas the \ion{Fe}{1} 630.15~nm line core is formed at $\log \tau =
-2.9$ in the quiet Sun, according to \cite{bal85}. These values are
consistent with the difference of approximately 270~km reported by
\cite{grec10} between the formation heights of the continuum and the
line core of \ion{Fe}{1} 630.15~nm. It is important to realize,
however, that the intensity observed at a given wavelength is 
affected by the conditions of a large fraction of the atmosphere
\citep[e.g.,][and references therein]{cab05}. Thus, any formation 
height actually represents a broad range of optical depths.

To investigate the magnetic properties of UDs we use simple line
parameters derived from the observed Stokes $Q$, $U$, and $V$
profiles. These are the mean linear polarization (LP) degree, 
\begin{equation}
{\rm LP} = \frac{\int^{\lambda _{\rm r}}_{\lambda _{\rm b}} 
[Q^2(\lambda) + U^2(\lambda)]^{1/2}/I(\lambda) \, 
{\rm d} \lambda}{\int^{\lambda _{{\rm r}}}_{\lambda _{{\rm b}}} {\rm d} \lambda},
\end{equation}
the mean circular polarization (CP) degree,
\begin{equation}
{\rm CP} = \frac{ \int^{\lambda _{\rm r}}_{\lambda _{\rm b}} 
[|V(\lambda)|/I(\lambda)] \, {\rm d} \lambda}{\int^{\lambda _{{\rm r}}}_{\lambda _{{\rm b}}} 
{\rm d} \lambda},
\end{equation}
and the net circular polarization (NCP), 
\begin{equation}
{\rm NCP} = \int_{\lambda_{\rm b}}^{\lambda_{\rm r}} V(\lambda) \, 
{\rm d} \lambda/I_{\rm c}.
\end{equation}
Here, $\lambda _{\rm r}$ and $\lambda_{\rm b}$ represent the limits of
integration over the line, and $I_{\rm c}$ is the continuum
intensity. The NCP accounts for the asymmetries in the Stokes $V$
profile, which are due to gradients in the magnetic field and the
line-of-sight (LOS) velocity \citep{illing75,auer78}.

In addition, we have determined the longitudinal component of 
the magnetic field using the center-of-gravity (COG) method
\citep{rees79,landi04}. The LOS component of the field is 
obtained through the relation 
\begin{equation}
B_{\rm LOS}=\frac{(\lambda_{+}-\lambda_{-})/2.0}{4.67\times10^{-13} \, 
g_{\rm L}\,\lambda_{0}^2},
\end{equation}
where $g_{\rm L}$ is the Land\'e factor, $\lambda_{0}$ is the central
wavelength of the line in \AA\/, and $\lambda_{\pm}$ are the centroids of
the right and left circularly polarized line components ($I \pm V$),
calculated as
\begin{equation}
\lambda_{\pm}=\frac{\int \lambda [I_{\rm c}-(I\pm V)] \, {\rm d} 
\lambda}{\int [I_{\rm c}-(I\pm V)] \, {\rm d} \lambda}
\end{equation}
\citep{uit03}. The COG method is not affected by saturation effects
in the strong field regime.

\begin{figure*}[t]
\centering
\vspace*{.30cm}
\resizebox{.28\hsize}{!}{\includegraphics{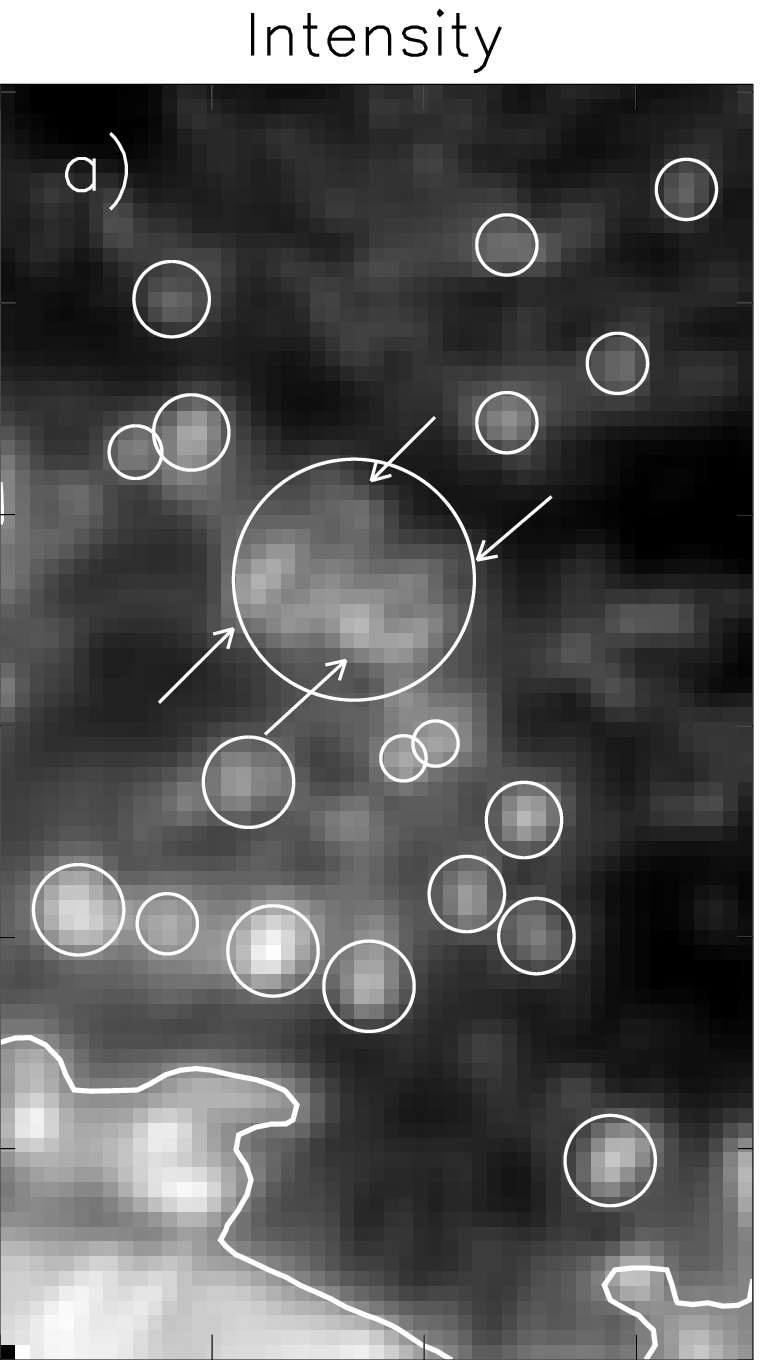}}
\resizebox{.28\hsize}{!}{\includegraphics{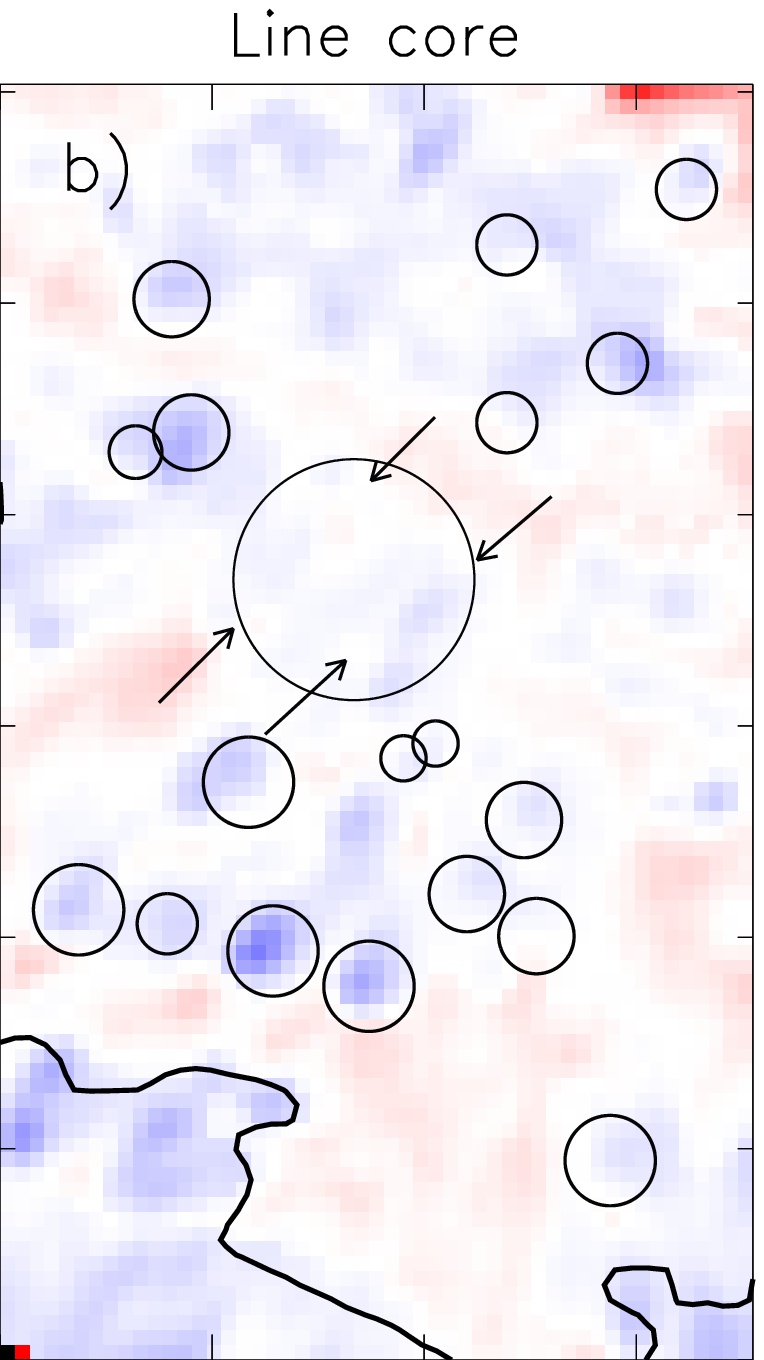}}
\resizebox{.28\hsize}{!}{\includegraphics{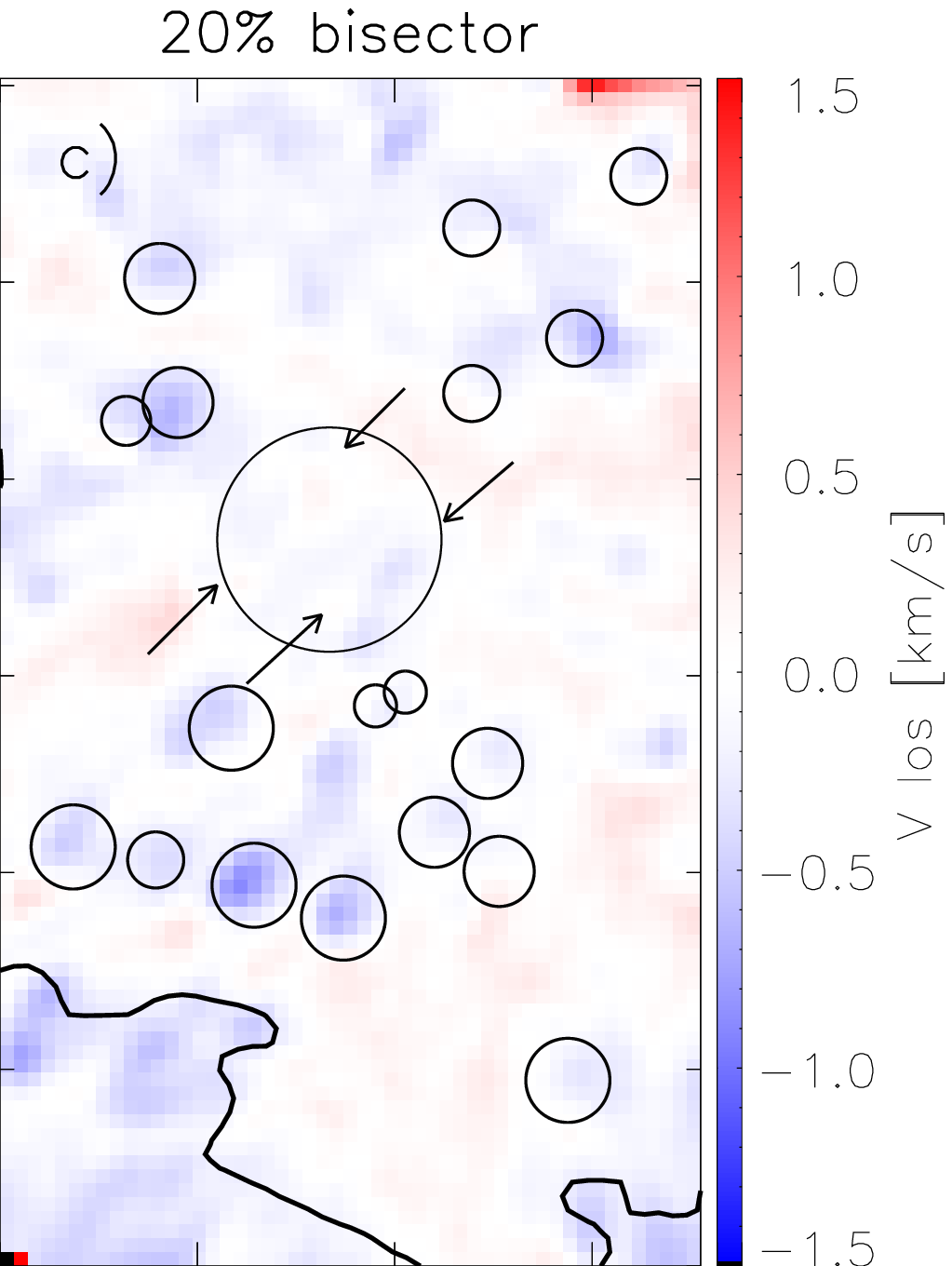}}\\
\vspace{1.25cm}
\resizebox{.28\hsize}{!}{\includegraphics{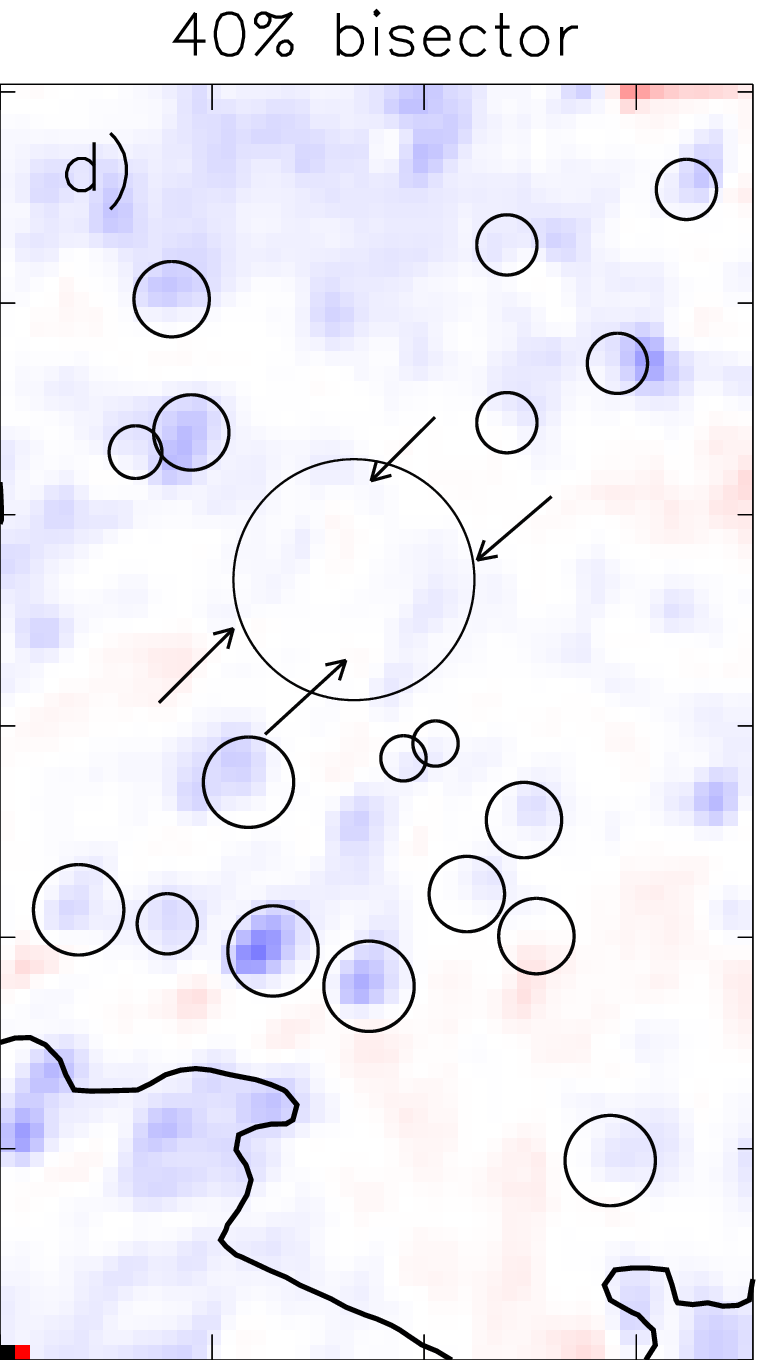}}
\resizebox{.28\hsize}{!}{\includegraphics{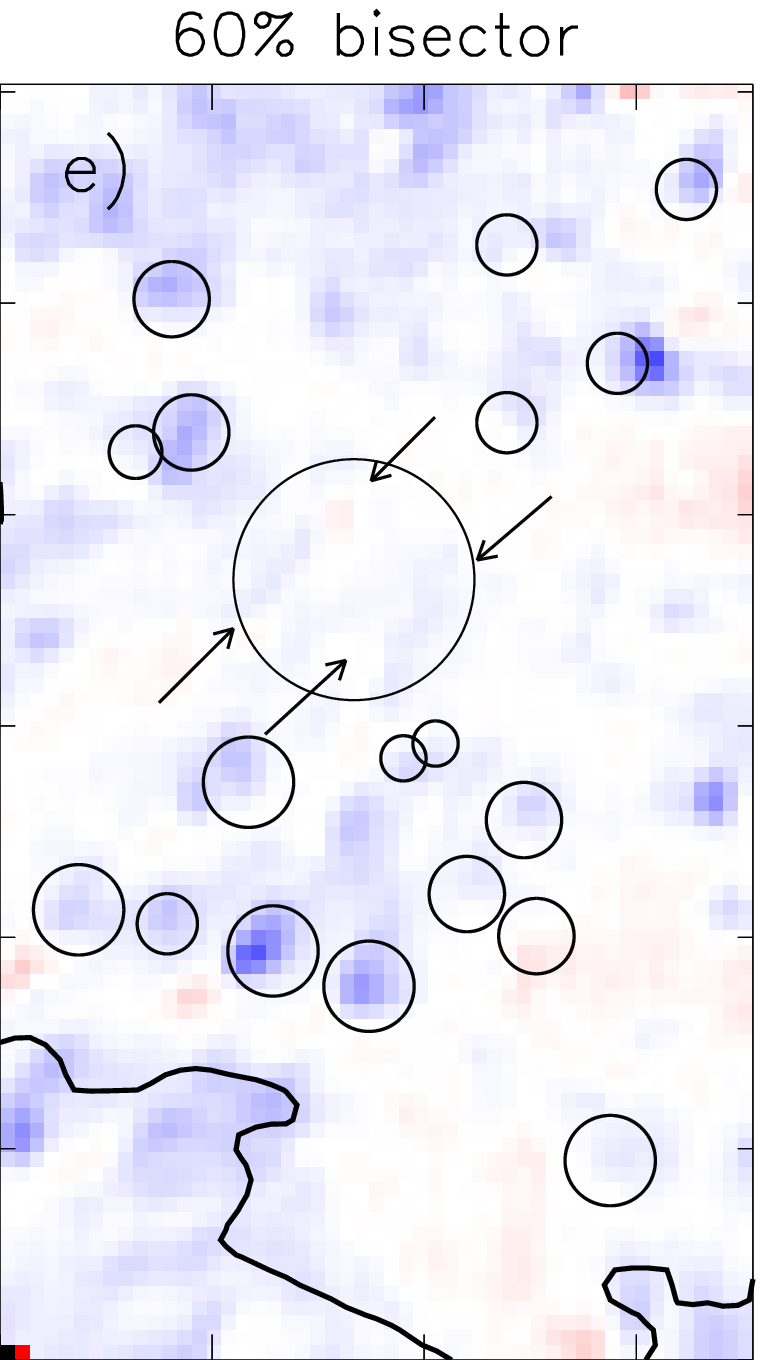}}
\resizebox{.28\hsize}{!}{\includegraphics{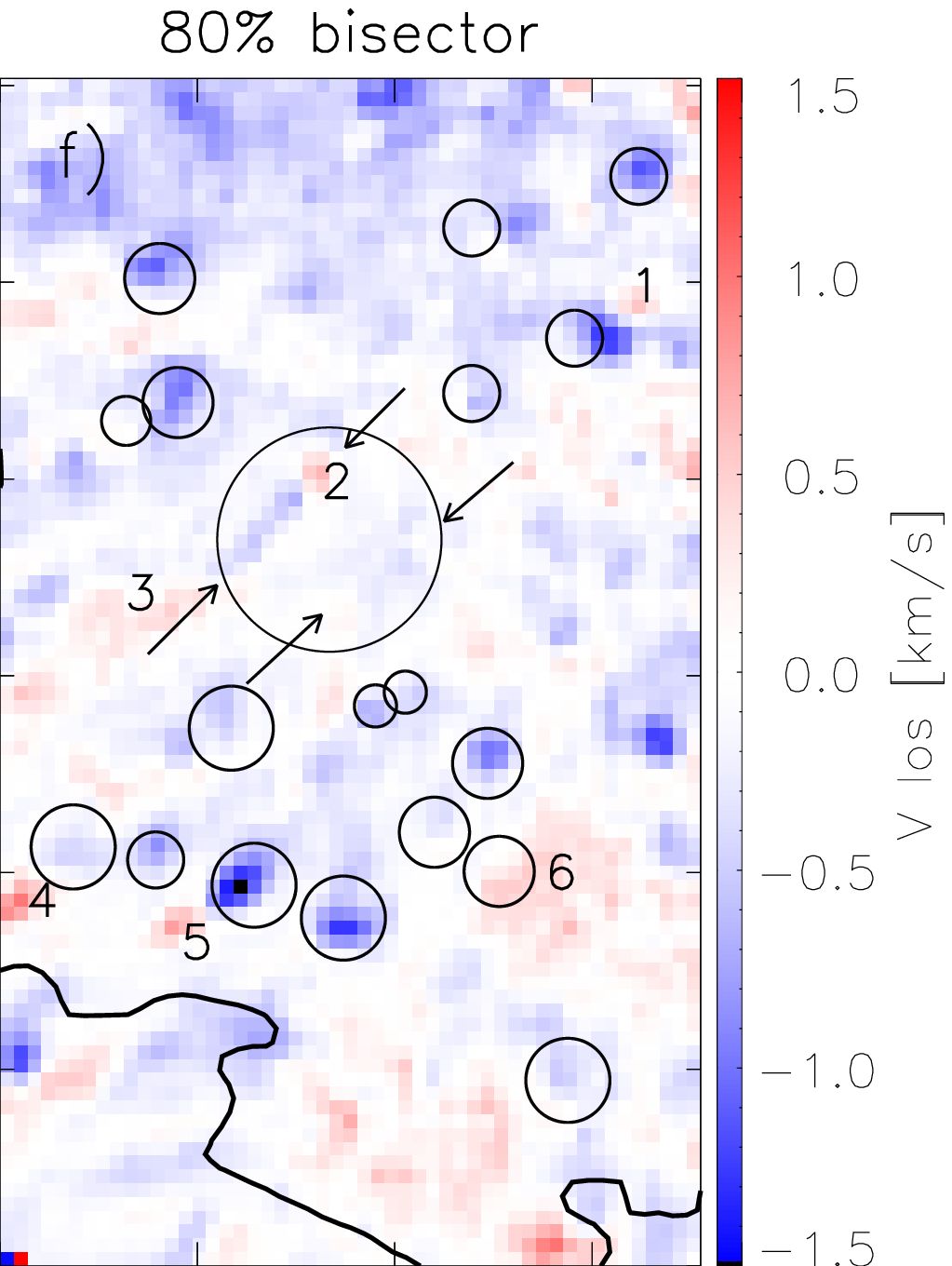}}
\vspace{.9cm}
\caption{LOS velocities in an area of the pore with high density 
of UDs (see the rectangle in Figure~\ref{fig1}). Distances are
measured in arcsec. The contour lines indicate the edge of the pore.
(a) Continuum intensity image. (b)-(f) Velocity maps at intensity 
levels from 0\% to 80\%. The circles enclose selected UDs and the
numbers refer to the downflow patches considered in Table~\ref{table1}.
\label{fig2}}
\vspace*{.3cm}
\end{figure*}

The inclination $\gamma$ of the magnetic field vector with respect 
to the LOS is determined as
\begin{equation}
\label{incl}
\gamma=\arccos\frac{\sqrt{1+x^2}-1}{x}, 
\end{equation}
where 
\begin{equation}
x=\frac{V}{\sqrt{Q^2+U^2}}. 
\end{equation}
This expression is valid for fully split lines
\citep{landi04}. We apply Equation~(\ref{incl}) to six
wavelength samples around the peaks of the Stokes $V$ profile (three
in the red lobe and another three in the blue lobe) and take the mean.
Once $B_{\rm LOS}$ and $\gamma$ are known, the magnetic field strength
$B$ is obtained as
\begin{equation}
B = B_{\rm LOS} / | \cos \, \gamma |.
\end{equation}

To reduce the noise, the line parameters LP, CP, and NCP, as well 
as the magnetic field parameters $B_{\rm LOS}$, $\gamma$, and $B$, 
have been computed separately for the two lines and then averaged.

\begin{deluxetable*}{cccccccc}
\tablecolumns{8}
\tablewidth{13cm}
\tablecaption{{\rm Downflows Observed in Selected UDs \label{table1}}}
\tablehead{
\colhead{Patch} & \colhead{Size} &  \colhead{Distance} & \colhead{$v_{\rm LOS}$} & \colhead{Distance} & \colhead{$v_{\rm LOS}$} & \colhead{UD size}  & \colhead{$I_{\rm max}/I_{\rm QS}$}  \\ 
\colhead{}      & \colhead{ } & \colhead{ } & \colhead{(m\,s$^{-1}$)} & \colhead{ } & \colhead{(m s$^{-1}$)} & \colhead{ } & \colhead{}  
}
\startdata
1 & $0\farcs21 \times 0\farcs21$ & 0\farcs21 & 580 & 0\farcs14 & $-1150$ & $0\farcs21 \times 0\farcs14$  & 0.71 \\
2 & $0\farcs21 \times 0\farcs21$ & 0\farcs28 & 680 & 0\arcsec  & $-620$  & $0\farcs70 \times 0\farcs35$  & 0.94  \\
3 & $0\farcs70 \times 0\farcs21$ & 0\farcs14 & 440 & 0\arcsec  & $-620$  & $0\farcs70 \times 0\farcs35$  & 0.94  \\
4 & $0\farcs28 \times 0\farcs28$ & 0\farcs21 & 990 & 0\arcsec  & $-380$  & $0\farcs28 \times 0\farcs35$  & 1.07  \\
5 & $0\farcs28 \times 0\farcs14$ & 0\farcs28 & 650 & 0\farcs07 & $-1420$ & $0\farcs28 \times 0\farcs28$  & 1.19 \\
6 &                              & 0\arcsec  & 520 &           &         & $0\farcs21 \times 0\farcs21$  & 0.78
\enddata
\tablecomments{
Column 1: number of downflow patch in Figure~\ref{fig2}(f). Column 2:
size of downflow patch. Column 3: distance of downflow patch to UD
intensity maximum. Column 4: downflow velocity at the 80\% intensity 
level. Column 5: distance of upflow patch to UD intensity maximum. Column 6: 
upflow velocity at the 80\% intensity level. Column 7: size of UD. Column 8: 
UD intensity maximum, normalized to the average quiet Sun continuum intensity. 
\vspace*{.3cm}
} 
\end{deluxetable*}

\section{Results}
\label{results}

\subsection{Morphological properties}
\label{morph}

Figure~\ref{fig1}(a) displays a continuum image of the pore as
observed on 2008 June 12 at 10:33 UT. The direction to disk center is
indicated by the arrow. The pore covers an area of around $6\arcsec
\times 9\arcsec$ and shows a complex network of magnetoconvective
structures (UDs) of different sizes, shapes, and brightnesses. In
general bigger structures are found in the lower half of the
pore. Some of the observed UDs present a dark lane along their
axis. Two of them even resemble {\em coffee beans} because of their
conspicuous dark lanes; one is located at $(x,y)=(4\arcsec, 6\arcsec)$
and the other at $(x,y)=(5\arcsec,5\farcs5)$. These two specific UDs
have a size of 0\farcs7 and the smallest dark lane is 0\farcs14 in
width (probably unresolved), in agreement with \citet{rim08} and
\citet{sobpus09}. Smaller UDs without a dark lane have a
typical diameter of 0\farcs3 and minimum sizes of less than
0\farcs2.

\subsection{Velocity field}
\label{vel}

The main goal of this work is to obtain the velocity field of the UDs
present in the pore. Figure~\ref{fig1}(b) shows the LOS velocities
derived from the bisector at the 80\% intensity level. The gray scale
is clipped at $\pm1.5$~km\,s$^{-1}$ to emphasize the weaker
flows. Negative velocities indicate blueshifts. Given the heliocentric
angle of the observations and the fact that the magnetic field is
nearly vertical, blueshifts most probably represent upflows and
redshifts downflows.

The granulation pattern is clearly visible, while the pore shows a
complex network of small-scale flow elements. One of the striking
features of this map is the strong downflows seen at the very edge of
the pore, near $(x,y)=(7\arcsec, 9\arcsec)$. They will be explored in
a forthcoming publication. Most of the umbra has small velocities
close to zero, but one can detect the presence of many upflows and
some downflow patches inside the pore. These structures are located at
the position of the UDs observed in Figure~\ref{fig1}(a), with small
deviations that we will discuss later. The upflows reach up to $-1400$
m\,s$^{-1}$, while the donwflows show a more modest value of around
400--1000 m\,s$^{-1}$ at the 80$\%$ intensity level.

Figure~\ref{fig2} displays the velocity field within the boxed area of
Figure~\ref{fig1}. For convenience, panel (a) shows the corresponding
continuum image. The circles mark selected UDs, and are centered at
the position of maximum brightness. The arrows indicate the two UDs
with dark lanes. Panels (b)--(f) show the bisector velocities at
five intensity levels, from 0$\%$ to 80$\%$. In this way, we probe
different layers of the photosphere.

\begin{figure*}[t]
\vspace*{0.5cm}
\centering
\hspace{-0.6cm}
\resizebox{.17\hsize}{!}{\includegraphics{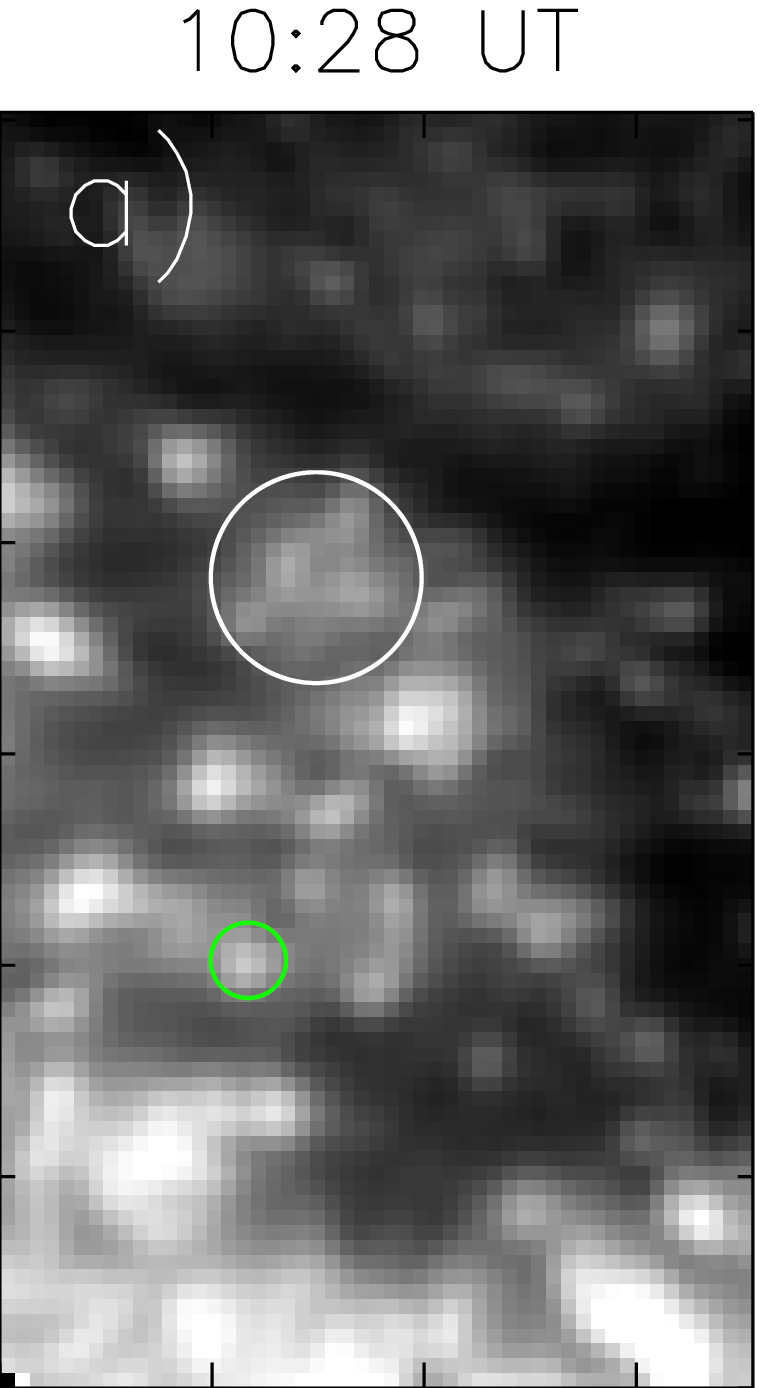}}
\hspace{-0.6cm}
\resizebox{.17\hsize}{!}{\includegraphics{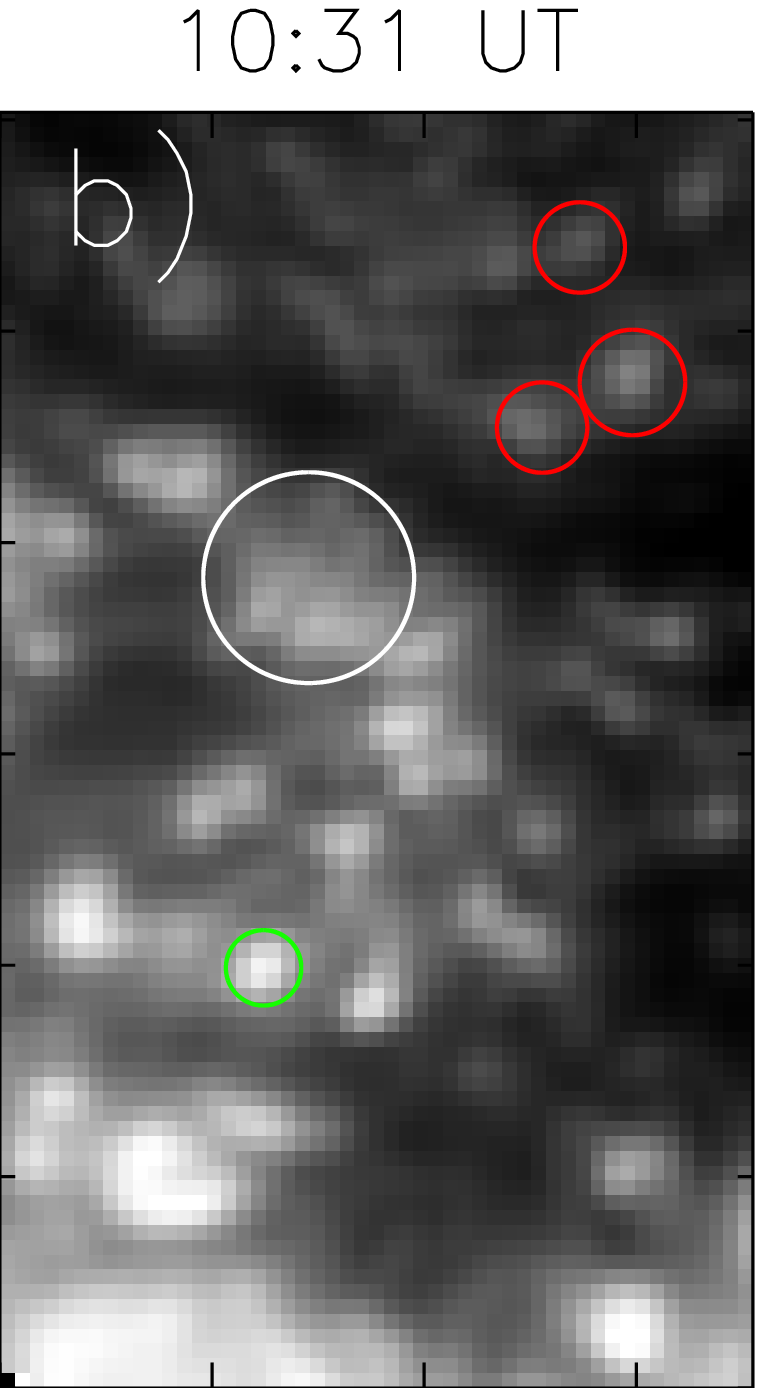}}
\hspace{-0.6cm}
\resizebox{.17\hsize}{!}{\includegraphics{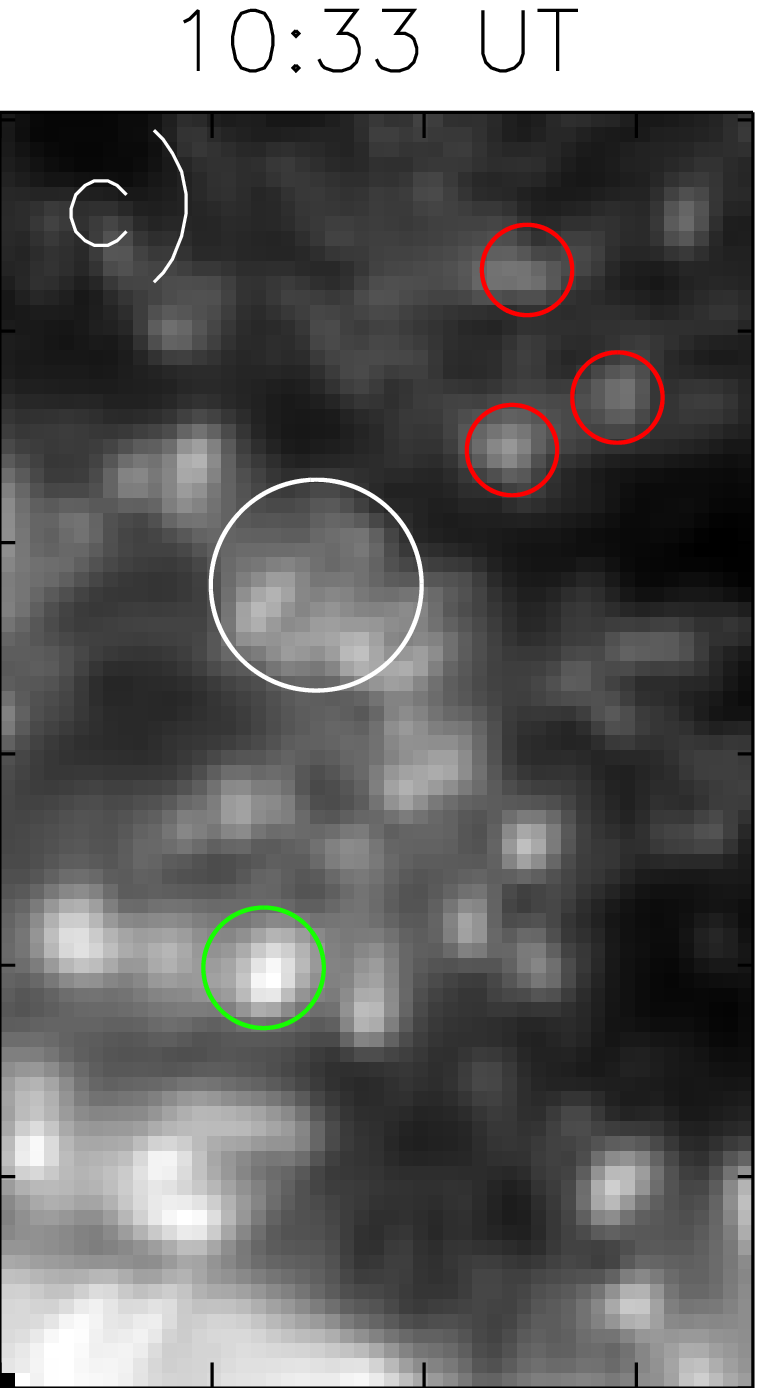}}
\hspace{-0.6cm}
\resizebox{.17\hsize}{!}{\includegraphics{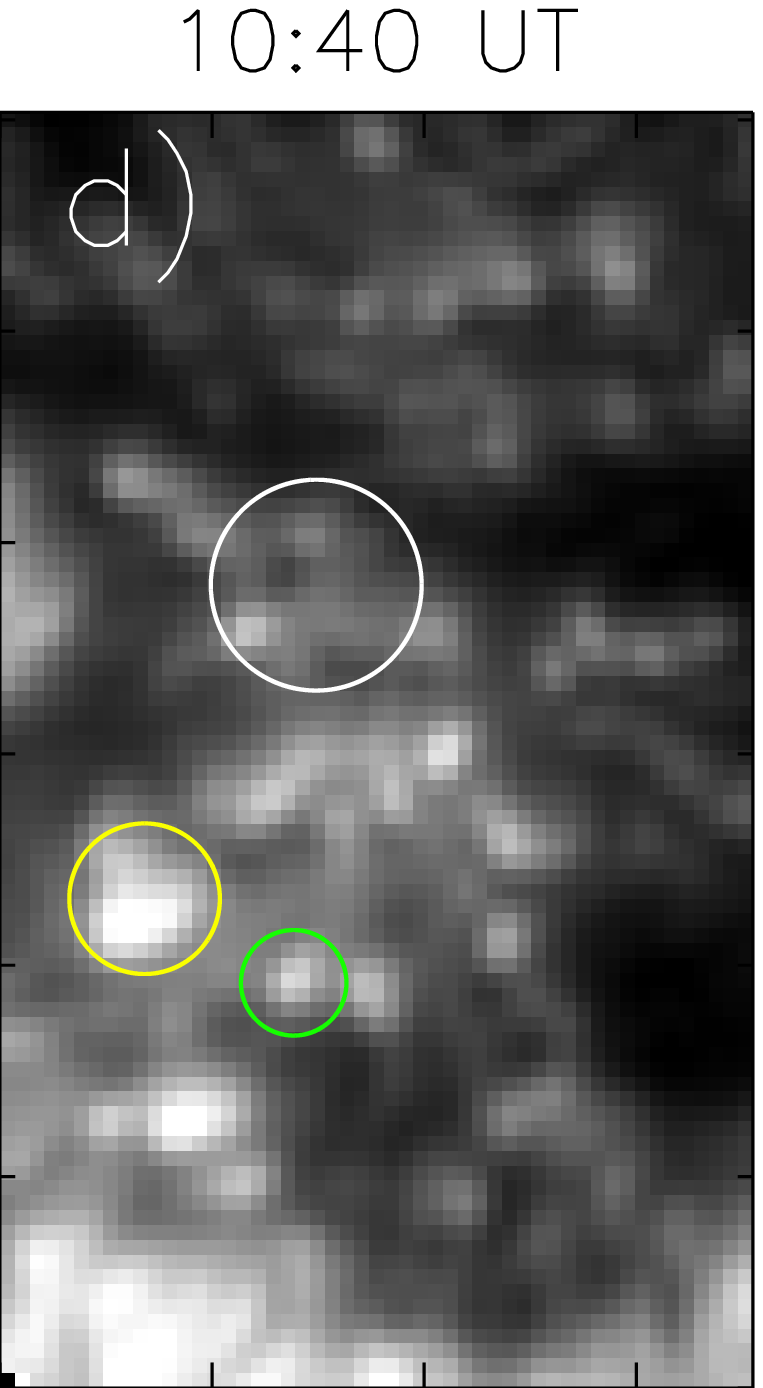}}
\hspace{-0.6cm}
\resizebox{.17\hsize}{!}{\includegraphics{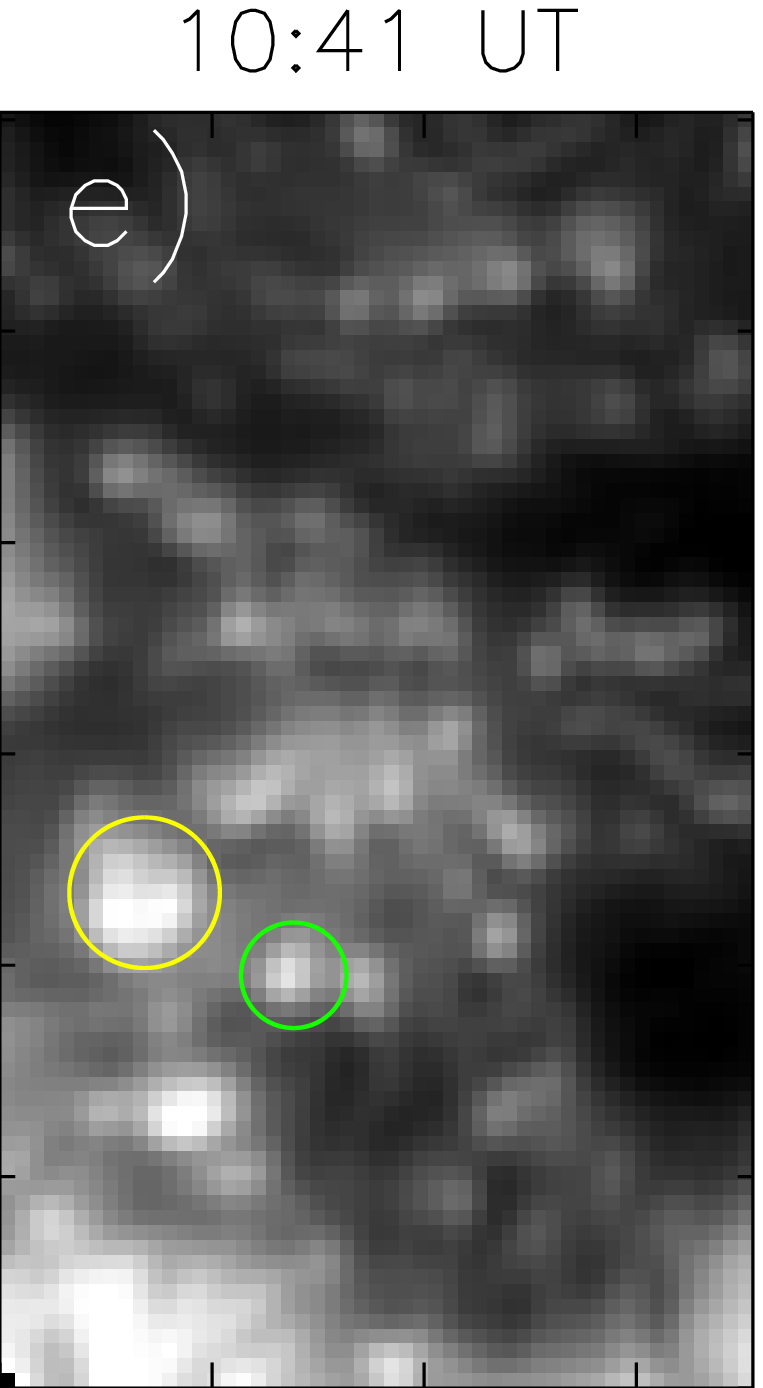}}
\hspace{-0.6cm}
\resizebox{.17\hsize}{!}{\includegraphics{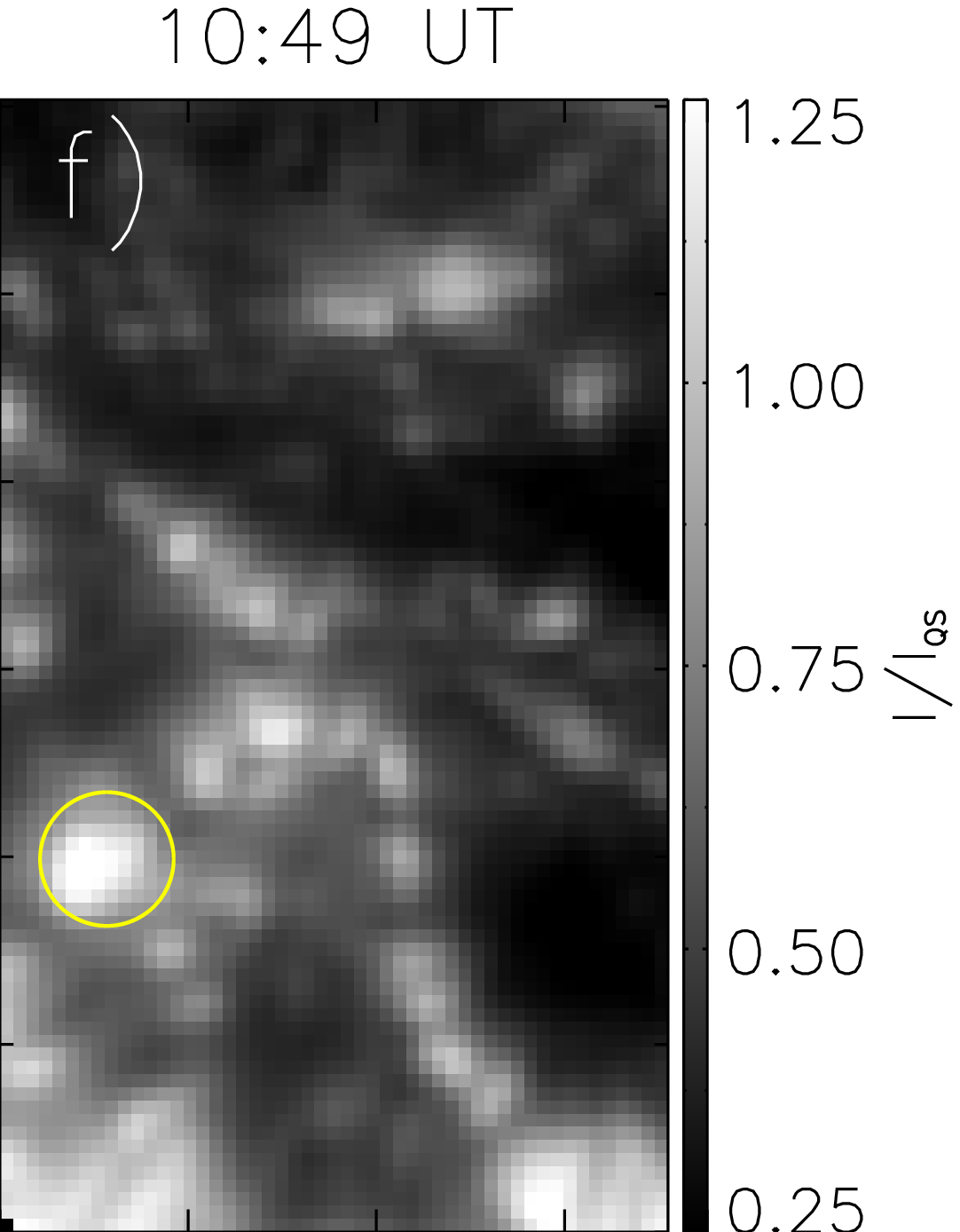}} \\
\hspace{-0.6cm}
\resizebox{.17\hsize}{!}{\includegraphics{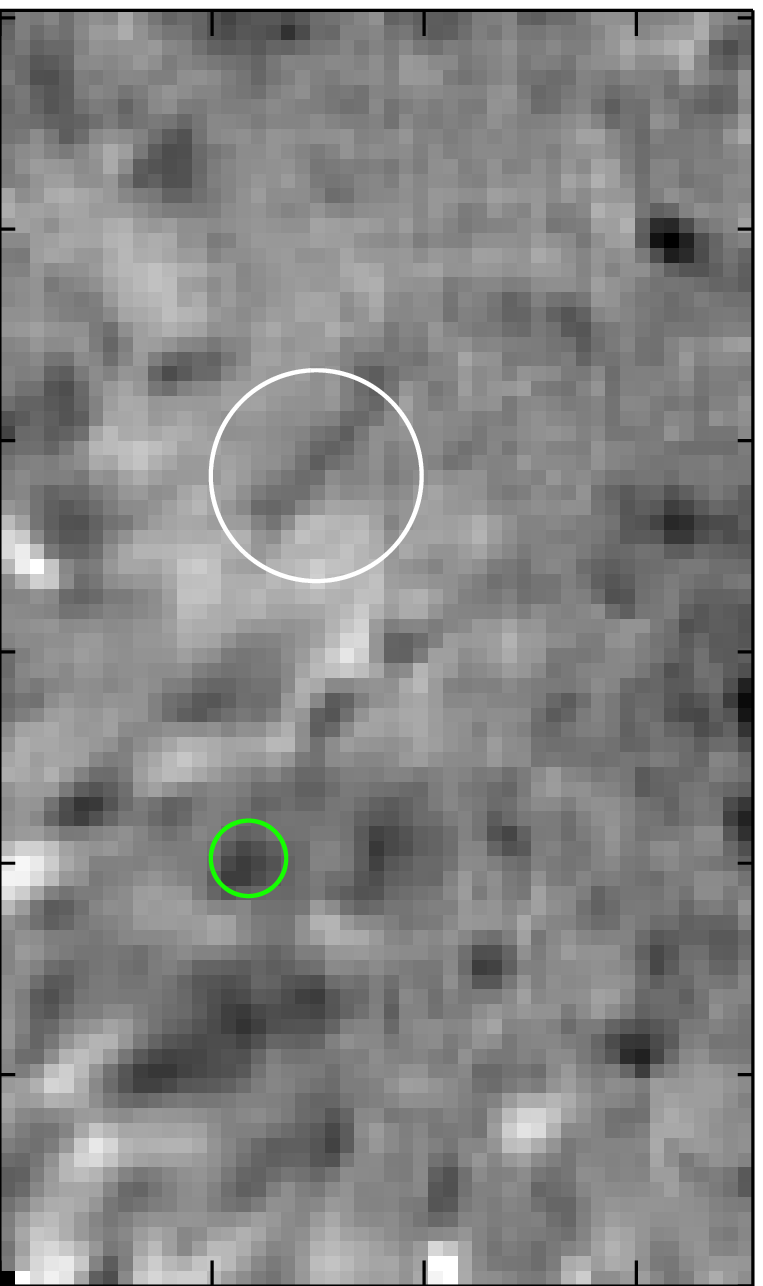}}
\hspace{-0.6cm}
\resizebox{.17\hsize}{!}{\includegraphics{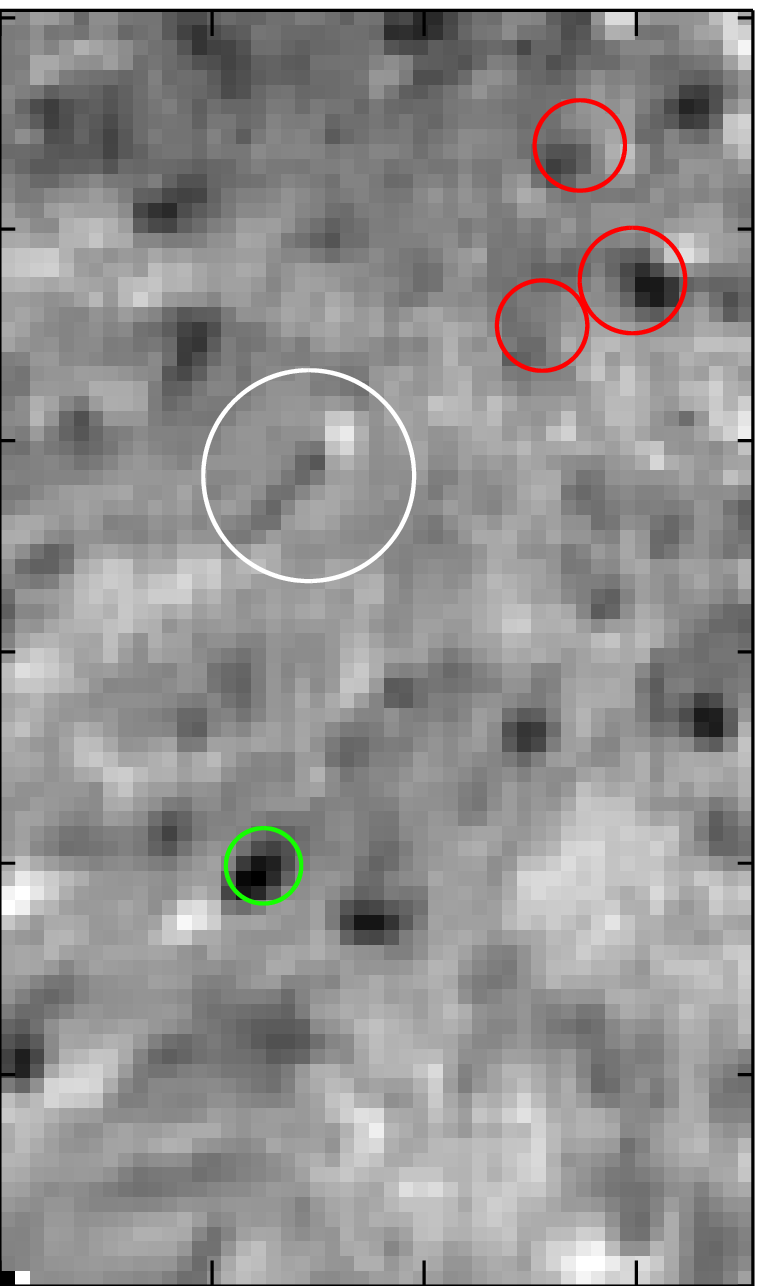}}
\hspace{-0.6cm}
\resizebox{.17\hsize}{!}{\includegraphics{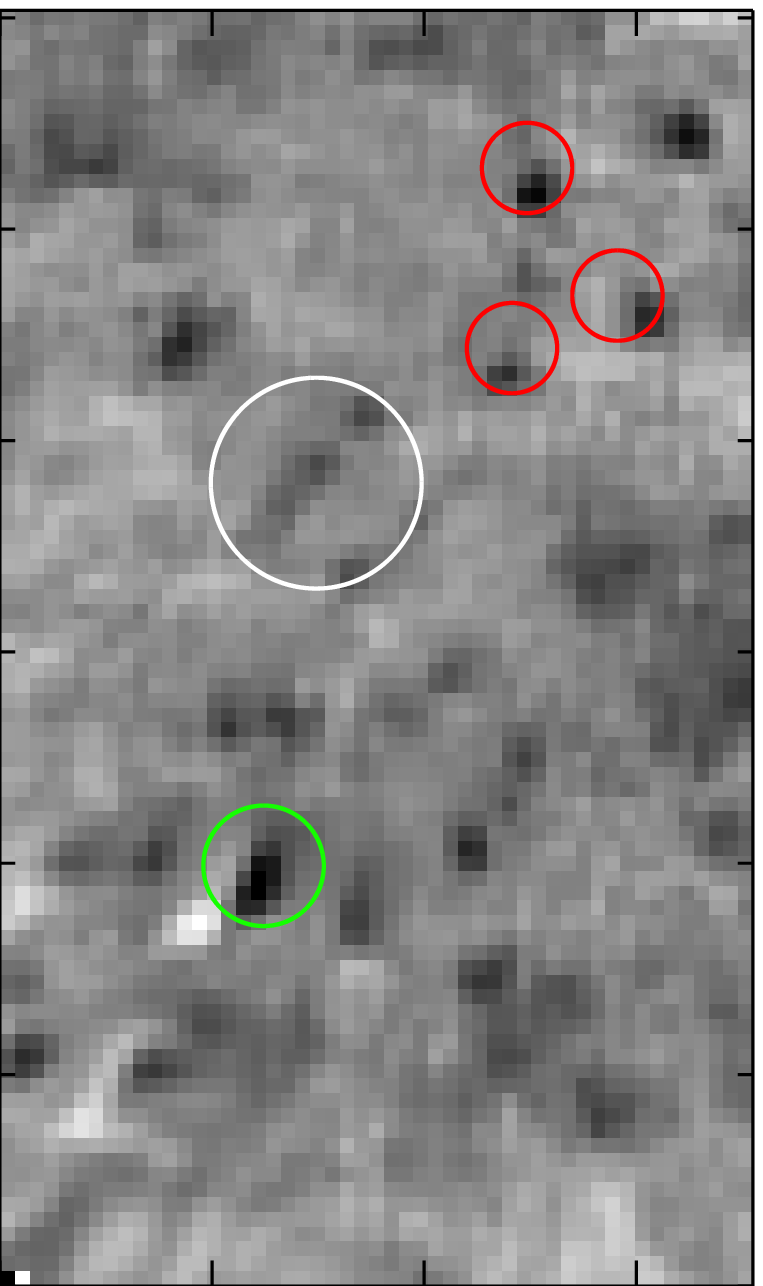}}
\hspace{-0.6cm}
\resizebox{.17\hsize}{!}{\includegraphics{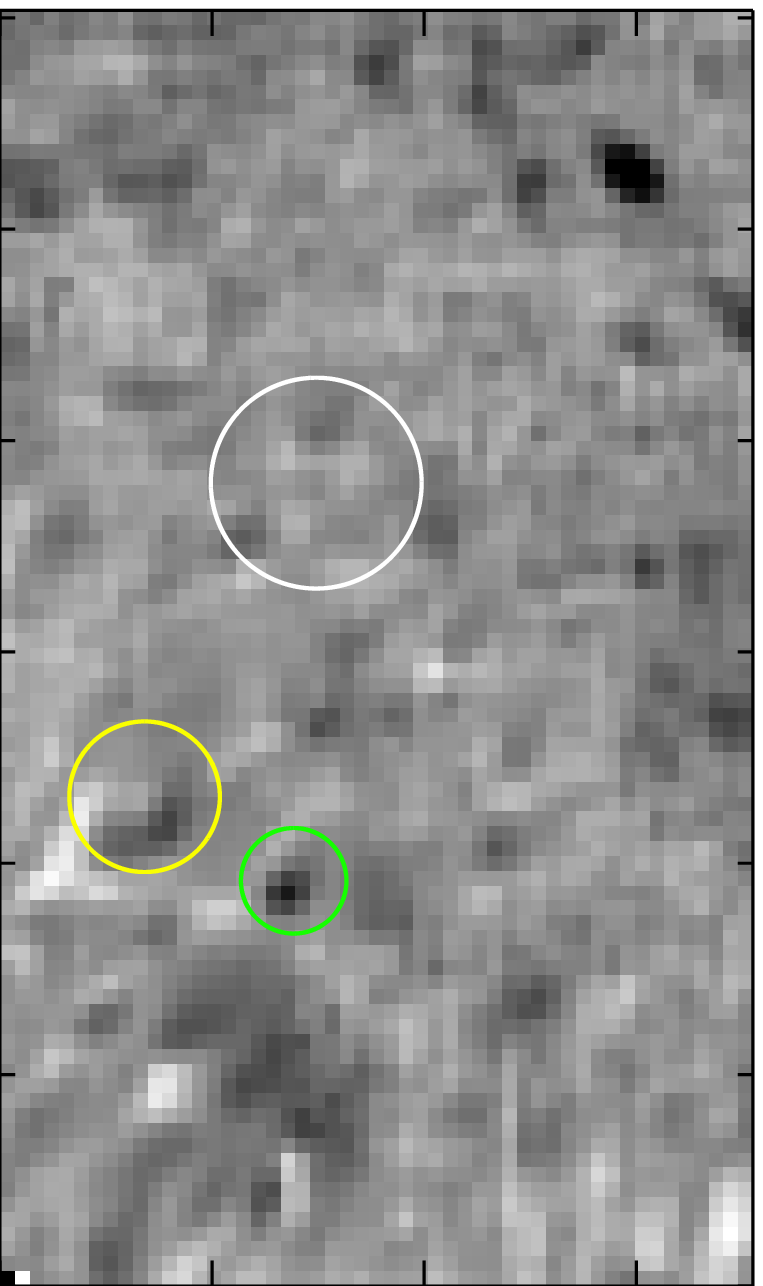}}
\hspace{-0.6cm}
\resizebox{.17\hsize}{!}{\includegraphics{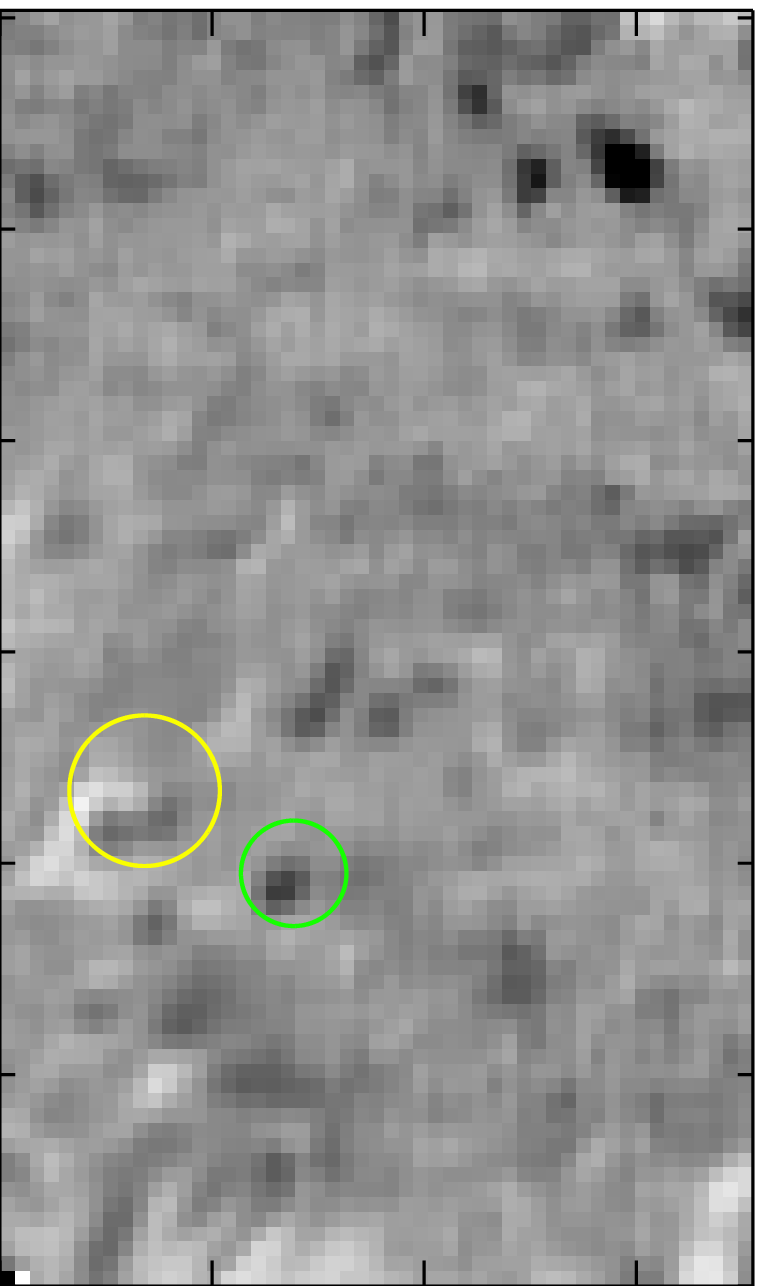}}
\hspace{-0.6cm}
\resizebox{.17\hsize}{!}{\includegraphics{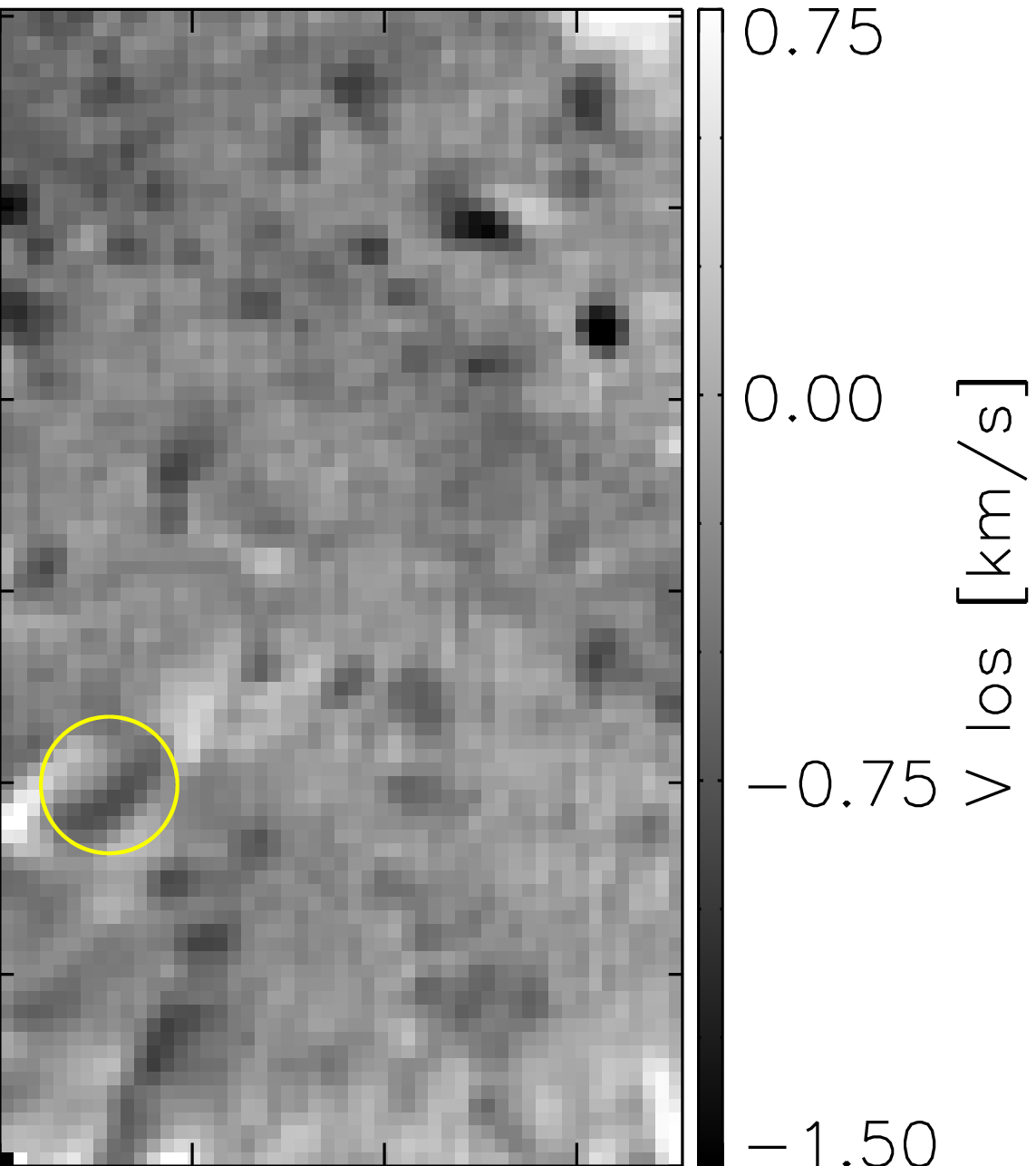}} \\
\hspace{-0.52cm}
\resizebox{.17\hsize}{!}{\includegraphics{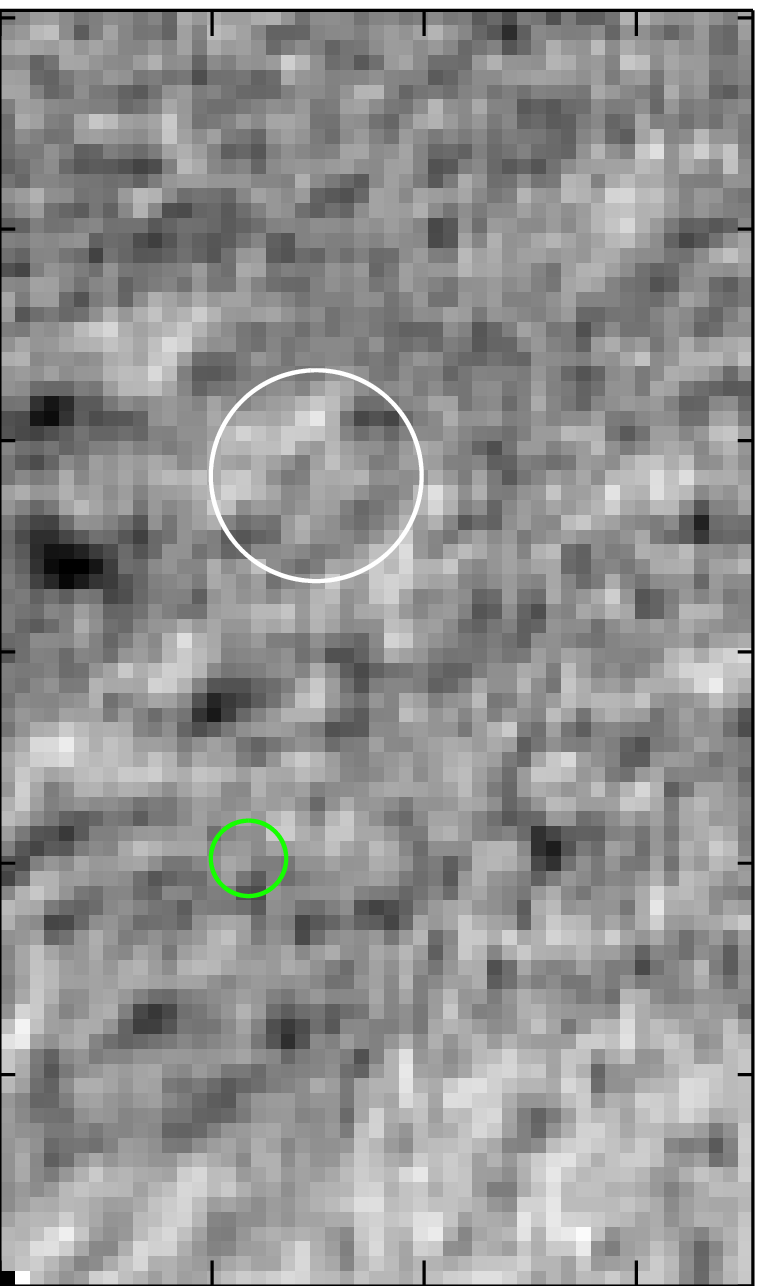}}
\hspace{-0.6cm}
\resizebox{.17\hsize}{!}{\includegraphics{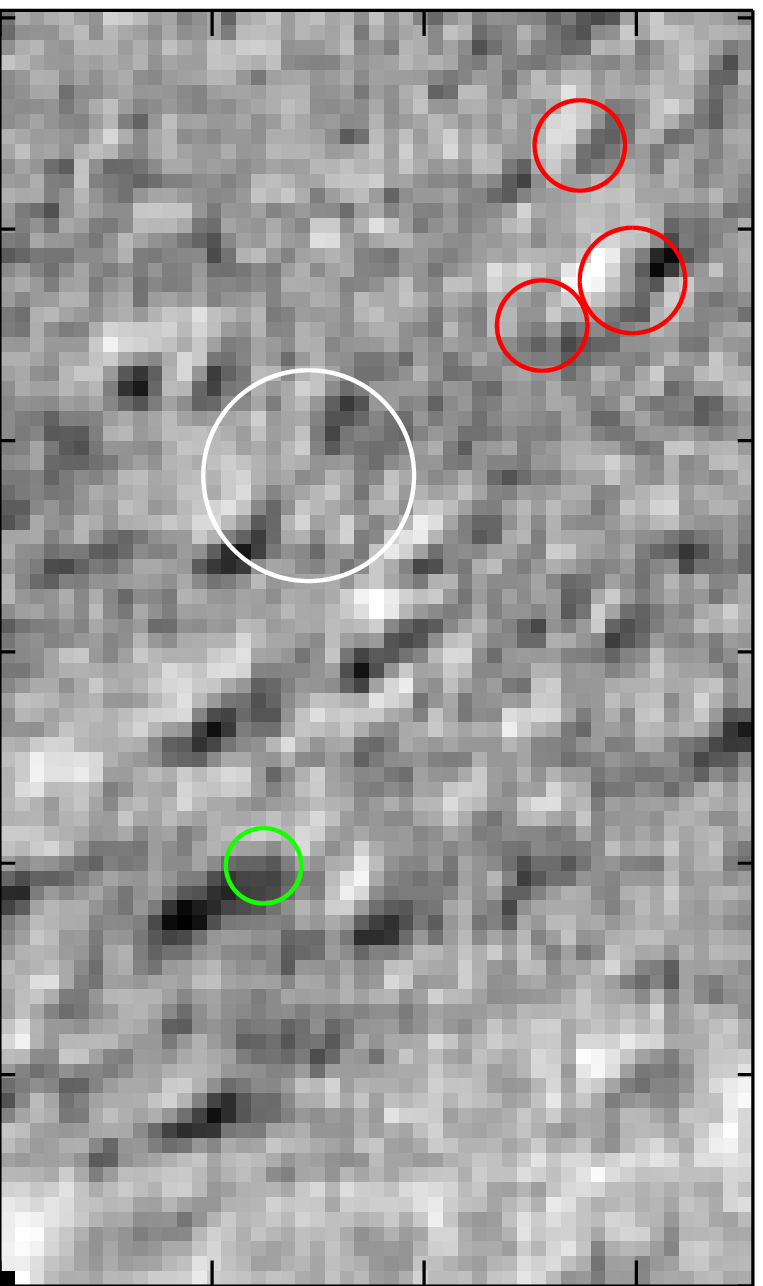}}
\hspace{-0.6cm}
\resizebox{.17\hsize}{!}{\includegraphics{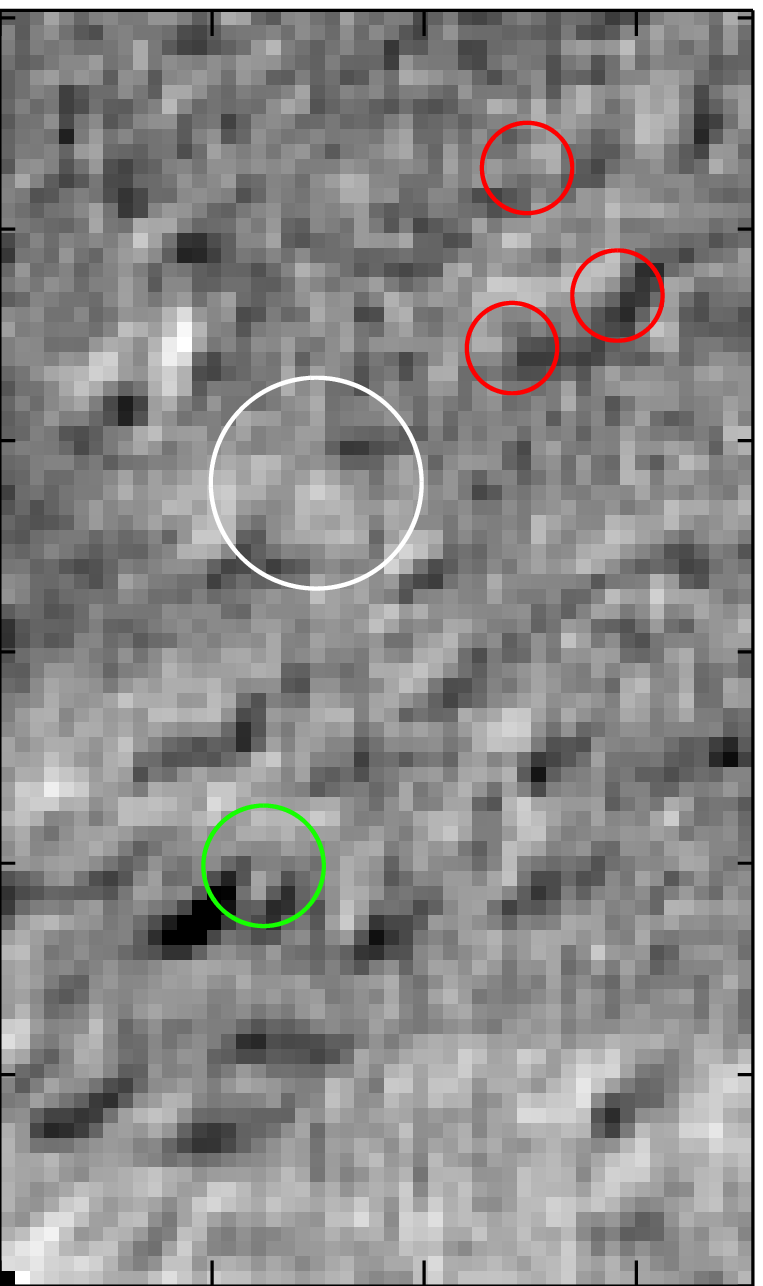}}
\hspace{-0.6cm}
\resizebox{.17\hsize}{!}{\includegraphics{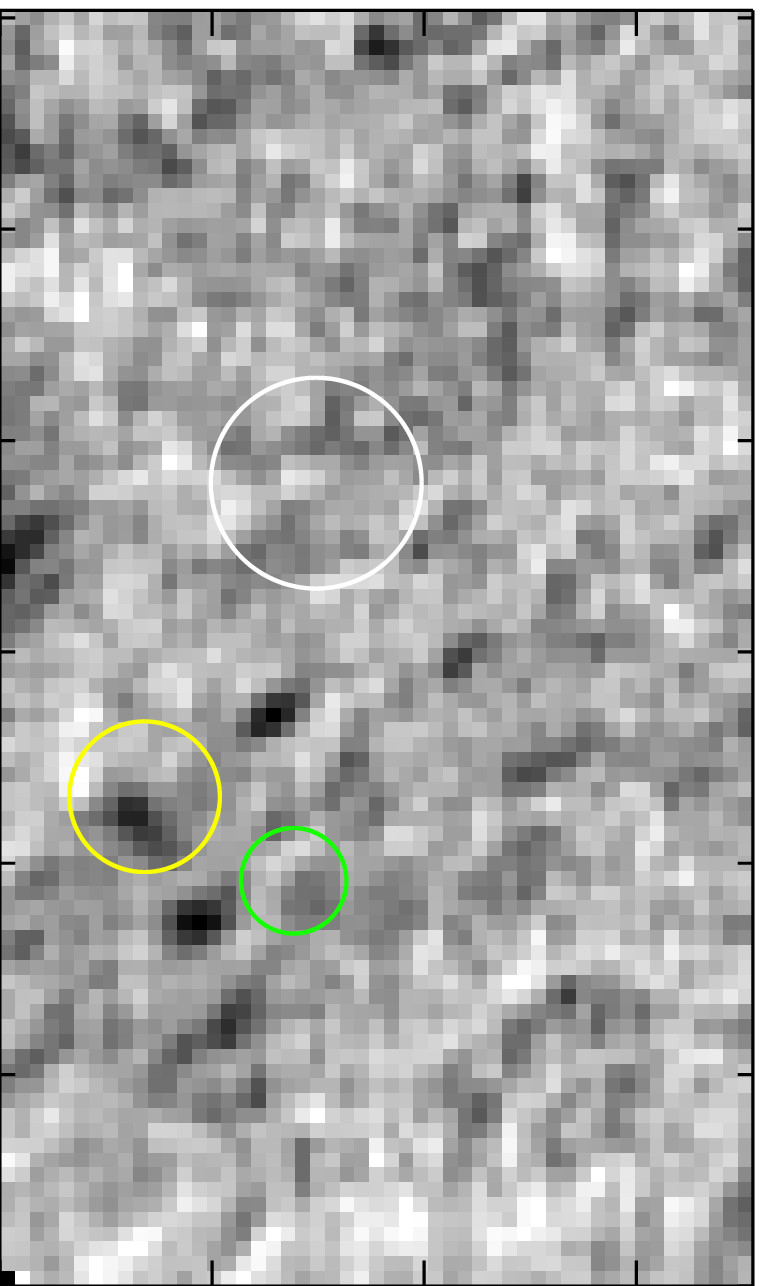}}
\hspace{-0.6cm}
\resizebox{.17\hsize}{!}{\includegraphics{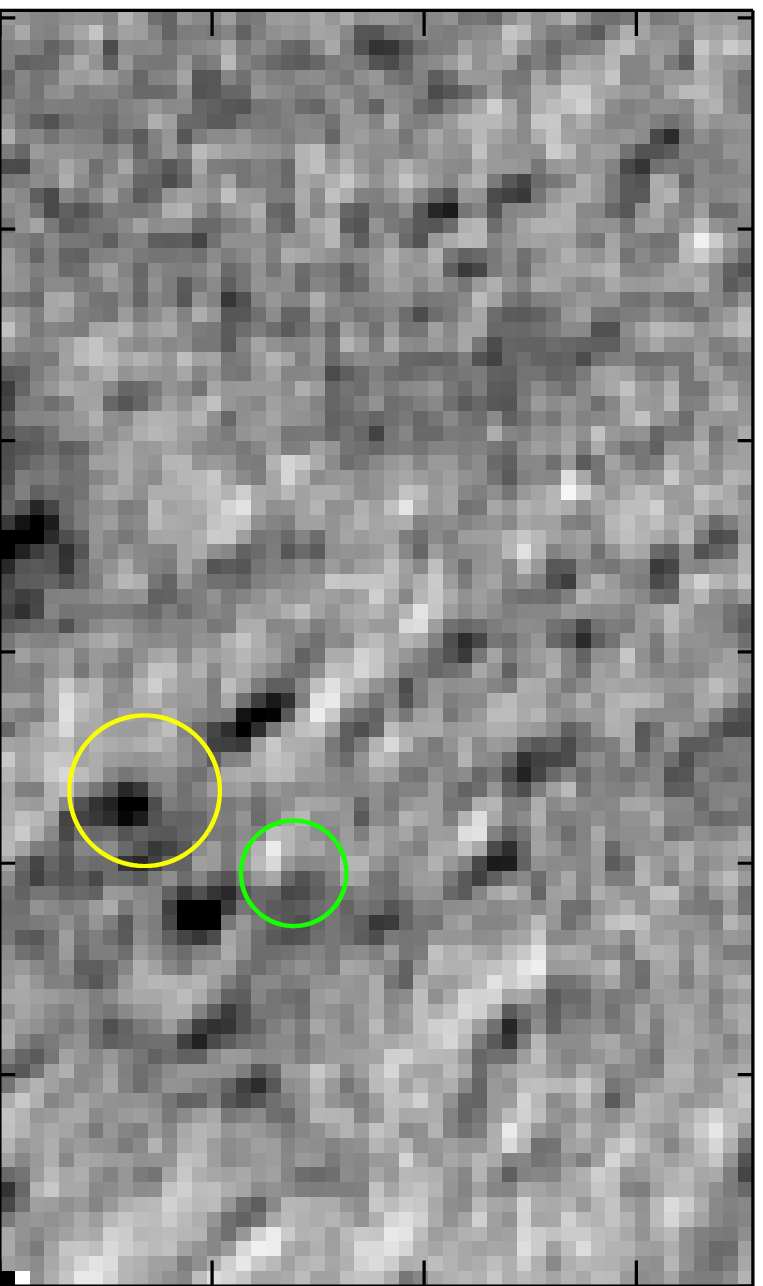}}
\hspace{-0.6cm}
\resizebox{.17\hsize}{!}{\includegraphics{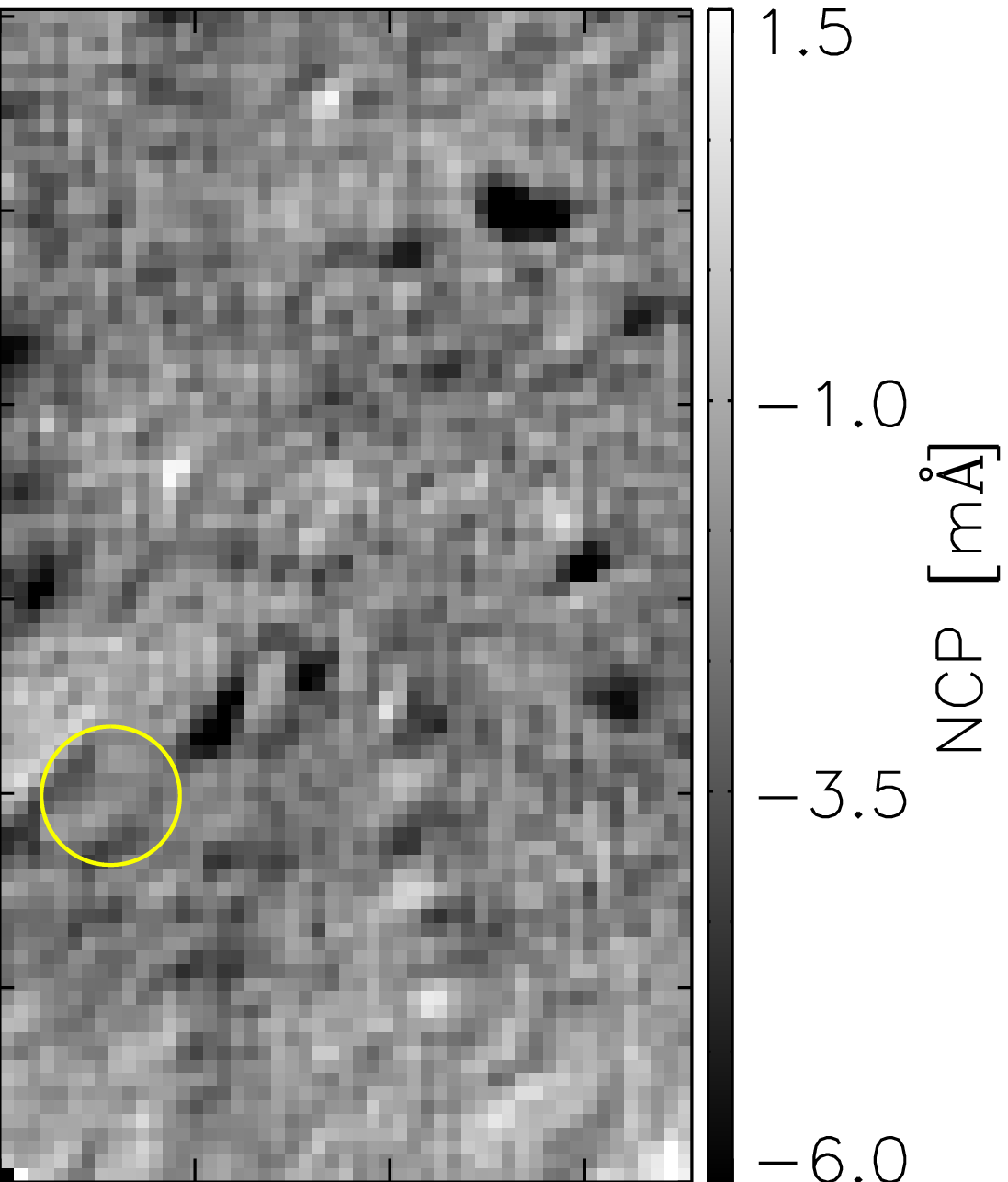}}
\vspace*{.7cm}
\caption{Temporal evolution of UDs in AR 10998. Panels (a)-(f) 
show six snapshots taken between 10:28 and 10:49 UT on 2008 June 12. Marks have been made every arcsecond. The upper panels show
continuum images zooming in the same area as Figure~\ref{fig2}. The
middle and lower panels display velocity maps derived from the $80\%$
bisector and NCP maps. The evolution of the UDs indicated by the
circles is discussed in the text.
\label{fig3}
\vspace*{.3cm}}
\end{figure*}

The bisectors close to the line core reveal small Doppler shifts. The
upflows and donwflows associated with the UDs become gradually
stronger toward the continuum, such that the larger velocities are
seen at the 80$\%$ intensity level [Figure~\ref{fig2}(f)]. Thus, the
flows occur preferentially in deep photospheric layers, and a gradient
of velocity with height appears to exist.  Such a gradient may partly
explain the difficulty of detecting flows in UDs; it is common to use
the line core for velocity determinations, but these measurements
probe too high layers already devoid of flows.

The majority of UDs show upflows, with maximum velocities of around
1400 m~s$^{-1}$ at the 80\% intensity level. The average size of the
upflow patches is around 0\farcs25, slightly smaller than the UDs
themselves. It is interesting to note that the location of the upflow
do not always exactly overlap with the position of maximum brightness
of the corresponding UD. Displacements of 0\farcs07--0\farcs14 (1-2
pixels)  are common at all intensity levels. This is
best seen in Figure~\ref{fig2}(f), where the photometric centers of
the UDs and the associated blue patches tend to be shifted, although
the flows always remain inside the circles representing the UDs.

Downflows are observed in some UDs, but almost exclusively at the
80$\%$ intensity level\footnote{The extended areas of weak redshifts
seen in the line core map coincide with the darkest parts of the
umbra, so they are not related to UDs. The redshifts could be produced
either by oscillations or by downward motions of the pore as a
whole.}. In those deep layers, they tend to appear at the edges of UDs exhibiting upflows. However, we note that not all UDs with
upflows have an associated downflow. Part of this may be due to their
transient nature (Section~\ref{temp}). In the case of the UDs showing
a dark lane, the upflows are cospatial with the dark lane and the
downflows occur at the endpoints.

The properties of some selected downflows are presented in
Table~\ref{table1}. The numbers in the first column identify the
corresponding patches in Figure~\ref{fig2}(f). All of them are related
to an adjacent upflow, except for patch 6 which does not seem to be
associated with any measurable upflow (this patch is embedded in a
large area of redshifts, so it may not be typical). The properties of
the upflows are given in Table~\ref{table1} along with those of the
parent UDs. The average size of the downflow patches is about
0\farcs25, but elongated areas of up to 0\farcs7 also exist. Since 
the downflows occur at the periphery of UDs, they can be shifted by as
much as 0\farcs28 (4 pixels) from the position of maximum
brightness. There seems to be a preferred direction of displacement
which coincides with the line of symmetry (i.e., the direction to disk
center), even though not in all cases, since patch 6 does not follow that
direction. The displacements might simply reflect an enhanced
visibility of the downflows near the line of symmetry due to
projection effects. The LOS velocities range from 400 m~s$^{-1}$ to almost 1000
m~s$^{-1}$ at the 80\% intensity level. We do not find any
relationship between the area of the downflow patch and its velocity
(patch 4, with the highest measured velocity, is not particularly big
or small), or between the area and the distance to the UD in continuum
intensity.

\subsection{Temporal evolution}
\label{temp}

We have also studied the temporal evolution of the UDs inside the
pore. Their morphological properties change rapidly with time. This is
demonstrated in Figure~\ref{fig3}, where six snapshots covering the
period from 10:28 UT to 10:49 UT are presented. The upper, middle, and
lower panels show continuum images, velocity maps derived from the
$80\%$ bisector, and NCP maps, respectively. The circles mark prototypical UDs whose
evolution we have followed in more detail.

The green circles trace the evolution of a small UD in the lower half
of the pore. The apparent size of this UD changes from a diameter of
0\farcs3 at 10:28 UT to 0\farcs5 at 10:33 UT, as seen in continuum
intensity. Not only does the diameter vary, but also the
velocity. This particular UD is interesting because it harbors both
strong upflows and downflows during part of its existence. It starts
showing an upflow of $-950$ m\,s$^{-1}$, but no downflow. At 10:31 UT
the upflows have increased to $-1530$ m\,s$^{-1}$, and a downflow patch
of around 700 m\,s$^{-1}$ has appeared next to the upflow patch. At
10:33 UT the upflows are the same while the downflows have increased
to 820 m\,s$^{-1}$. At 10:40 and 10:41 UT, the motions seem to slow
down, reaching velocities of $-950$ m\,s$^{-1}$ and 180 m\,s$^{-1}$,
respectively. The sizes of the upflow and downflow patches do not
change much in this particular example.

At 10:49 UT, a fast downflow of 950 m\,s$^{-1}$ becomes very prominent
in the lower left part of the image (yellow circles). This patch had
existed at least during three snapshots, but it is in the last one
when it shows the highest velocities. The downflows have the shape 
of a collar surrounding a large UD with upflows of $-780$ m\,s$^{-1}$.

The red circles pinpoint a group of three UDs in the upper part of the
FOV. They are well visible at 10:31 and 10:33 UT but have already
disappeared at 10:40 UT. At 10:31 UT, the rightmost UD shows an upflow
of $-1280$ m\,s$^{-1}$ and an adjacent downflow of 470
m\,s$^{-1}$. The other two UDs harbor upflows of $-950$ m\,s$^{-1}$ on
average. At 10:33 UT, the downflow patch has vanished and the upflows
have decreased to $-1150$ m\,s$^{-1}$.

\begin{figure*}
\centering
\resizebox{0.60\hsize}{!}{\includegraphics[bb= 0 -20 697 566]{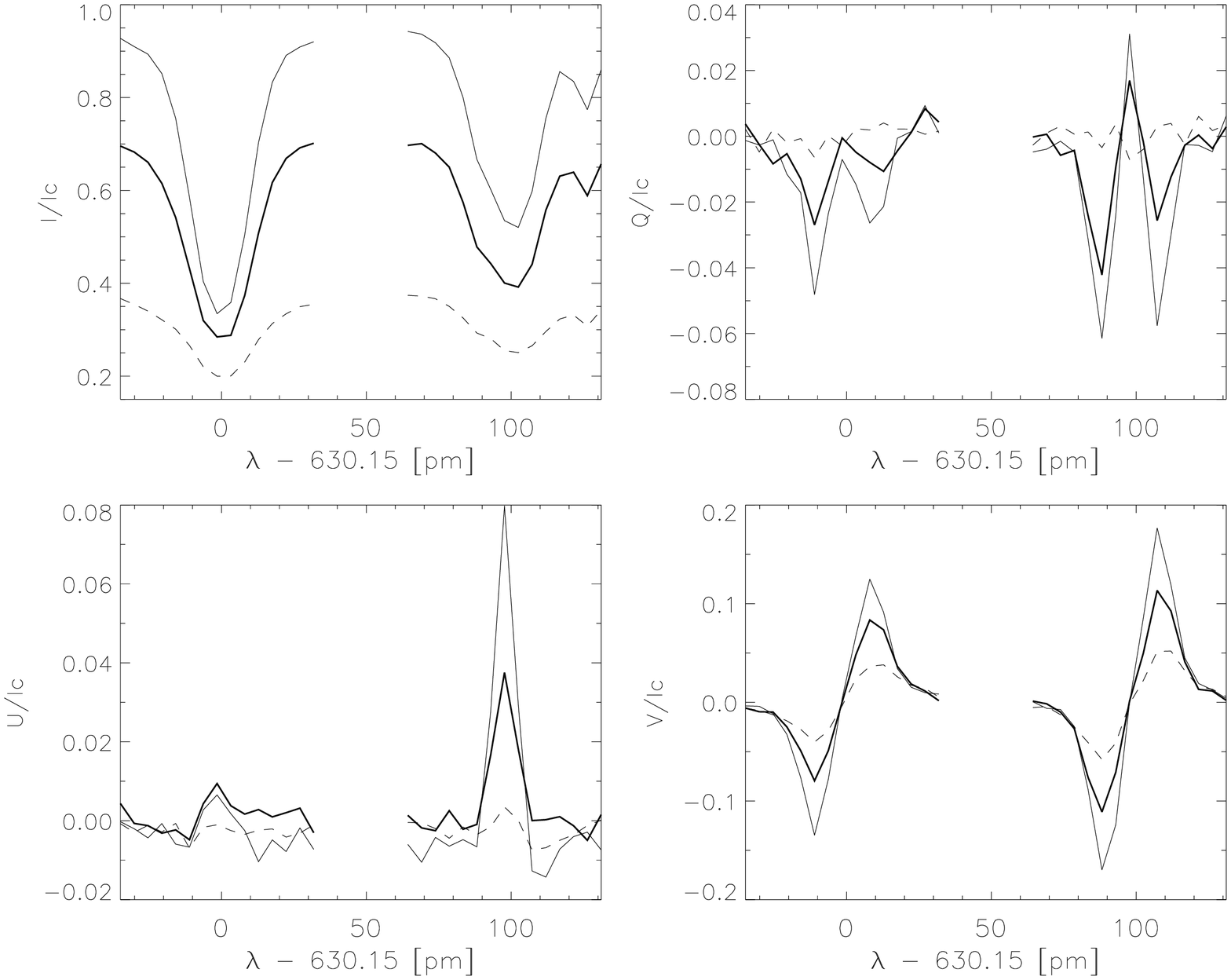}}
\caption{Stokes $I$, $Q$, $U$, and $V$ profiles emerging from 
a UD ({\em solid line}), its dark lane ({\em thick solid line}), 
and the adjacent umbra ({\em dashed line}). The profiles are
normalized to the quiet Sun continuum intensity.\label{fig4}}
\end{figure*}

Finally, the white circles highlight one of the UDs that contain a
dark lane. The dark lane appears in intensity at 10:31 UT and
disappears at 10:40 UT as the whole structure becomes blurry. In the
velocity maps, we can observe narrow, elongated upflows that coincide
with the position of the dark lane. Their velocities are $-670$
m\,s$^{-1}$ at 10:28 UT, $-730$ m\,s$^{-1}$ at 10:31 UT, and $-860$
m\,s$^{-1}$ at 10:33 UT. At 10:31 UT, downflow patches of $560$
m\,s$^{-1}$ appear at the two endpoints of the dark lane. One of the
patches is as small as 0\farcs14 in diameter. Their velocities quickly
decrease with time; 2 minutes later, at 10:33 UT, the downflows are
no longer visible.

All of these cases corroborate: (1) the ephemeral life of the UDs
themselves (around 10 minutes), (2) the ephemeral life of their
substructures (the dark lanes are well defined for only around 4 minutes),
and (3) the quick change in their velocities, with upflow/downflow
motions lasting only for a few minutes. In fact,
\citet{sobpus09} report that the substructures observed within UDs
(i.e., the dark lanes) vary with a typical time scale of about 3
minutes. According to \citet{sobpus09}, dark lanes disappear and reappear
during the evolution of central UDs. Our time sequence is relatively
short, and we have been unable to observe the reappearance of any dark
lane.

\subsection{Polarization signals and magnetic properties}

In Figure~\ref{fig4}, we present the four Stokes profiles of an UD, its
dark lane, and the surrounding umbra. The selected UD is the one
enclosed by the white circle in Figure~\ref{fig3}. With a continuum
approaching that of the quiet Sun, this UD is brighter than the
adjacent umbra by a factor of about 2.6. \cite{sobotka93} deduced that
the brightness of UDs is at most 3 times that of the background
umbra. Our observations appear to confirm this result, derived from a
careful analysis of high-resolution photometric measurements. A factor
of 2.6 is also in agreement with the average continuum ratio of 2.58
obtained from magnetoconvection simulations at 630~nm \citep{bhar10}. 
The dark lane, in turn, is observed to be about 25\% darker than 
the brightest part of the UD.

The UD and the dark lane show significantly larger Stokes $V$ signals
than the umbral background. \citet{hector04} pointed out that this
does not imply stronger magnetic fields, but is just a consequence of
a steeper source function gradient in hot structures. The UD and the
dark lane also show enhanced Stokes $Q$ and $U$ signals, suggesting
more inclined fields. Overall, the spectra displayed in
Figure~\ref{fig4} are very similar to those presented by
\cite{hector04}, except that these authors did not resolve the
contribution of the dark lane.

\begin{figure*}[t]
\centering
\hspace{0.40cm}
\resizebox{.29\hsize}{!}{\includegraphics[bb= 0 0 255 415]{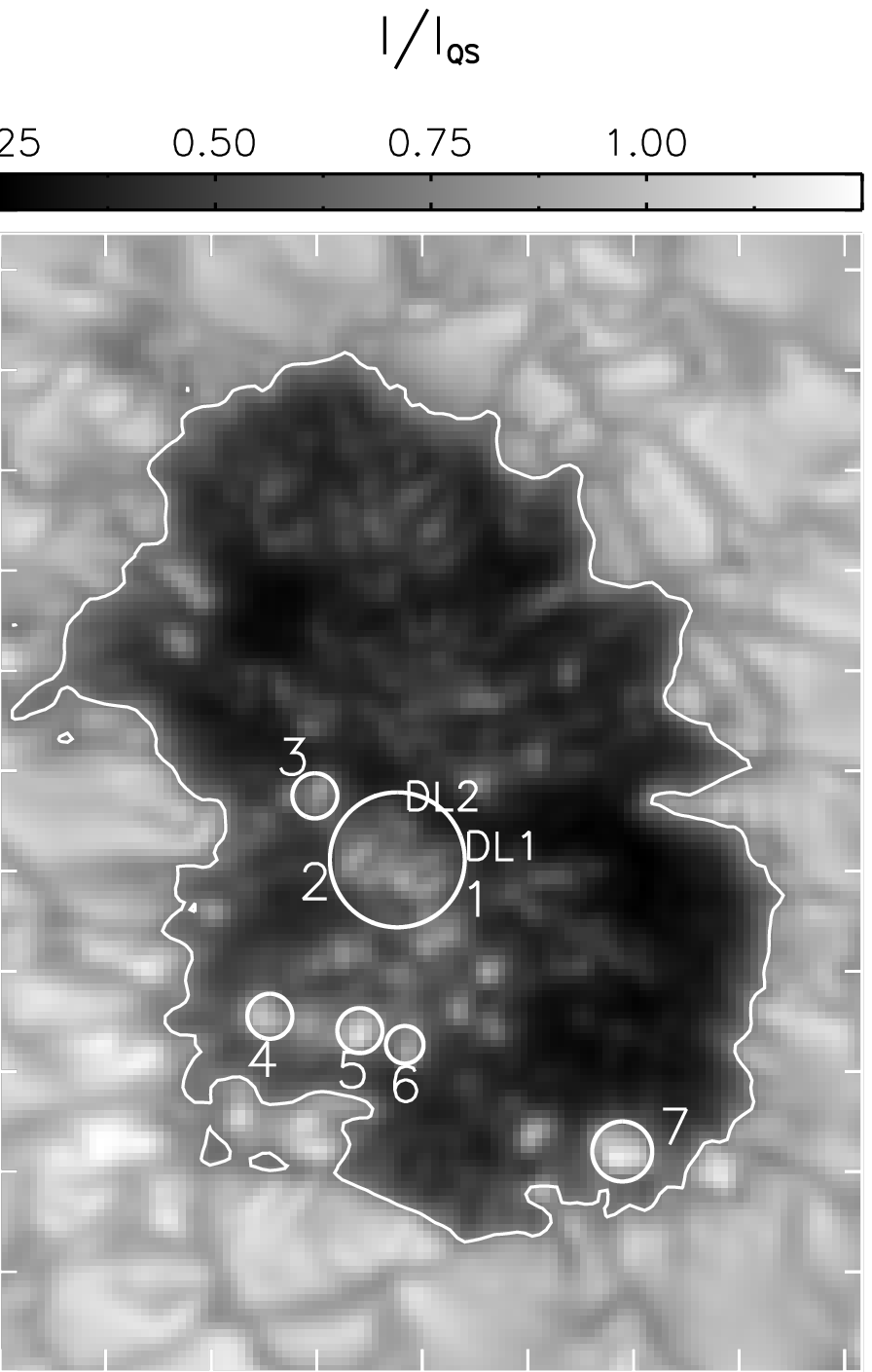}}
\hspace{0cm}
\resizebox{.29\hsize}{!}{\includegraphics[bb= 0 0 255 415]{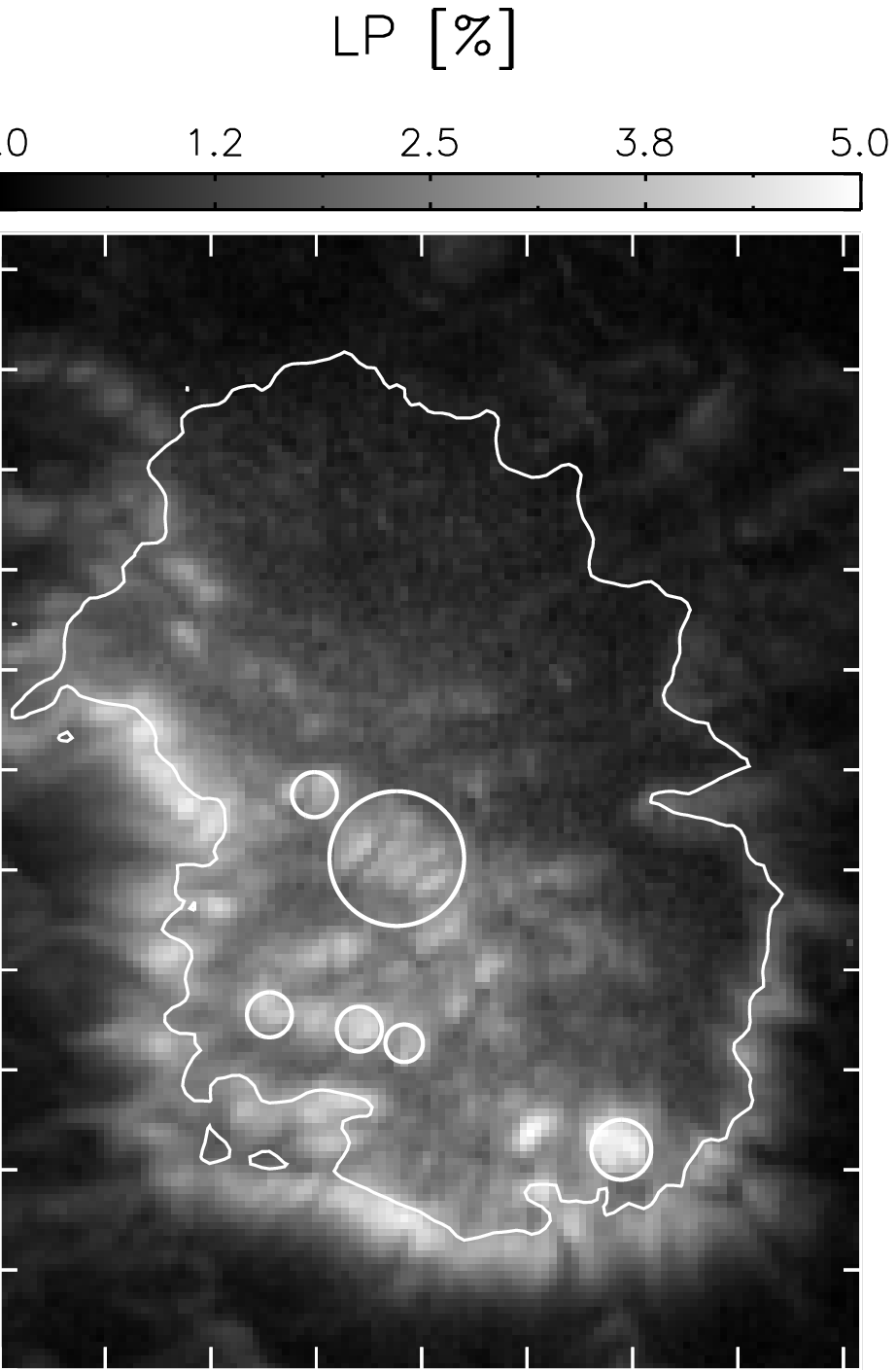}}
\hspace{0cm}
\resizebox{.29\hsize}{!}{\includegraphics[bb= 0 0 255 415]{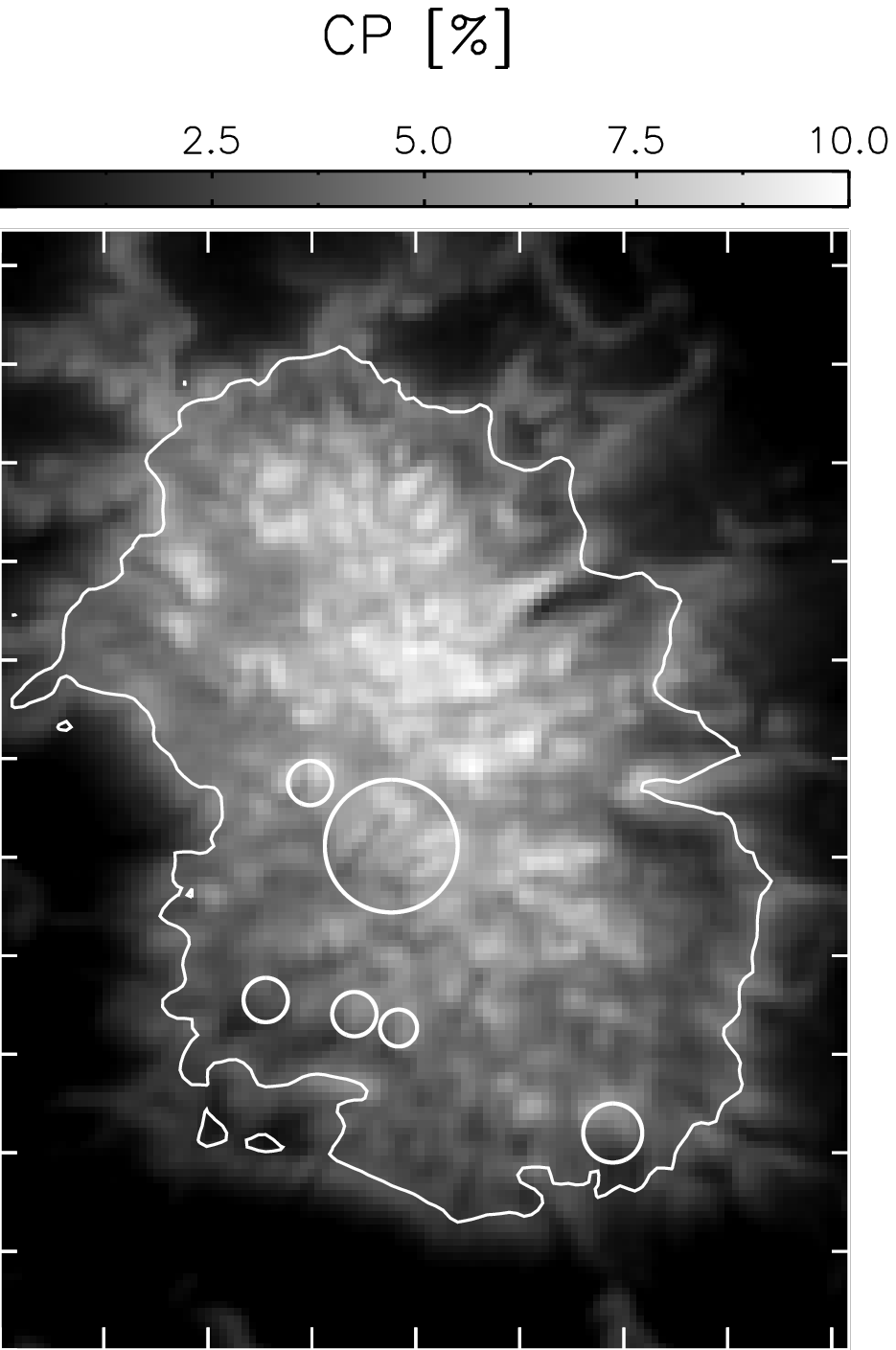}}\\
\hspace{0.48cm}
\resizebox{.29\hsize}{!}{\includegraphics[bb= 0 0 255 360]{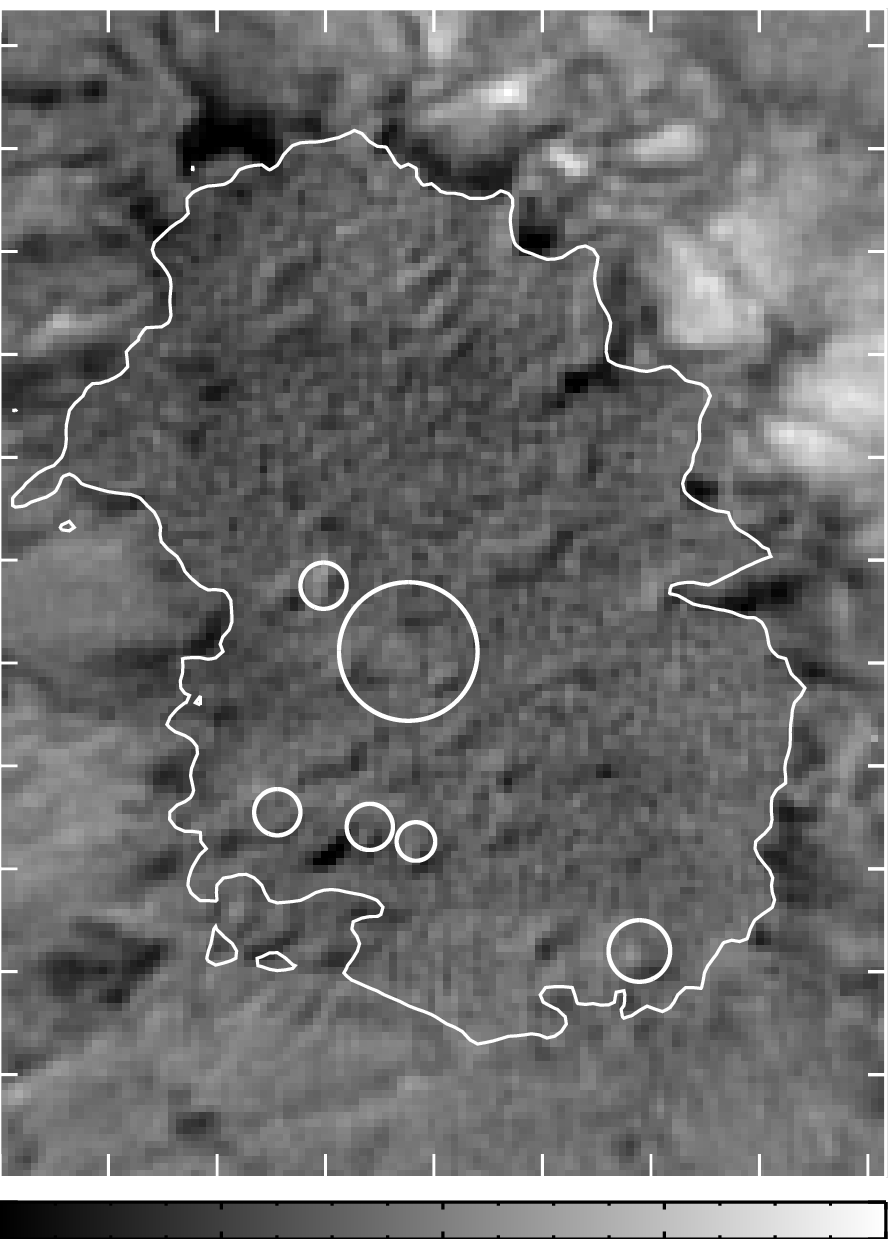}}
\hspace{0cm}
\resizebox{.29\hsize}{!}{\includegraphics[bb= 0 0 255 360]{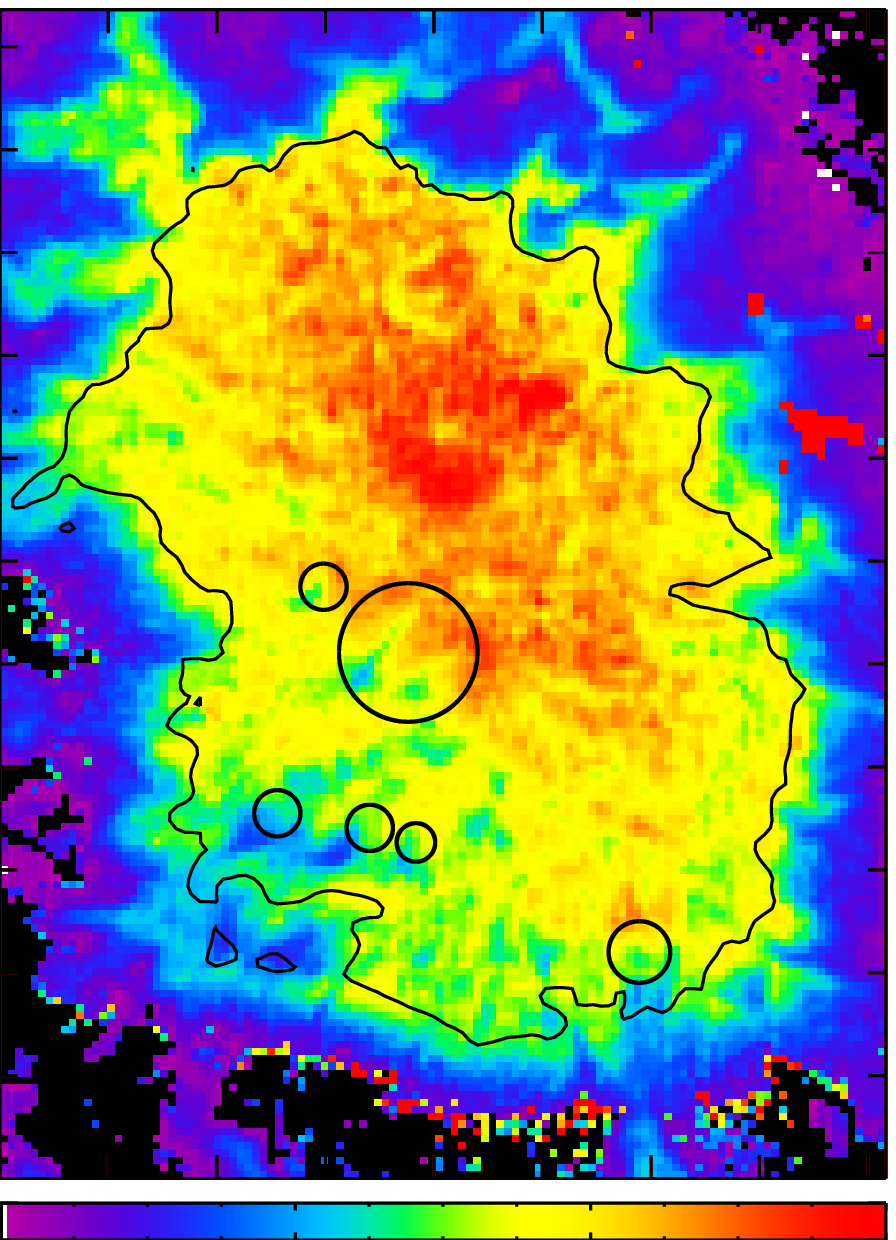}}
\hspace{0cm}
\resizebox{.29\hsize}{!}{\includegraphics[bb= 0 0 255 360]{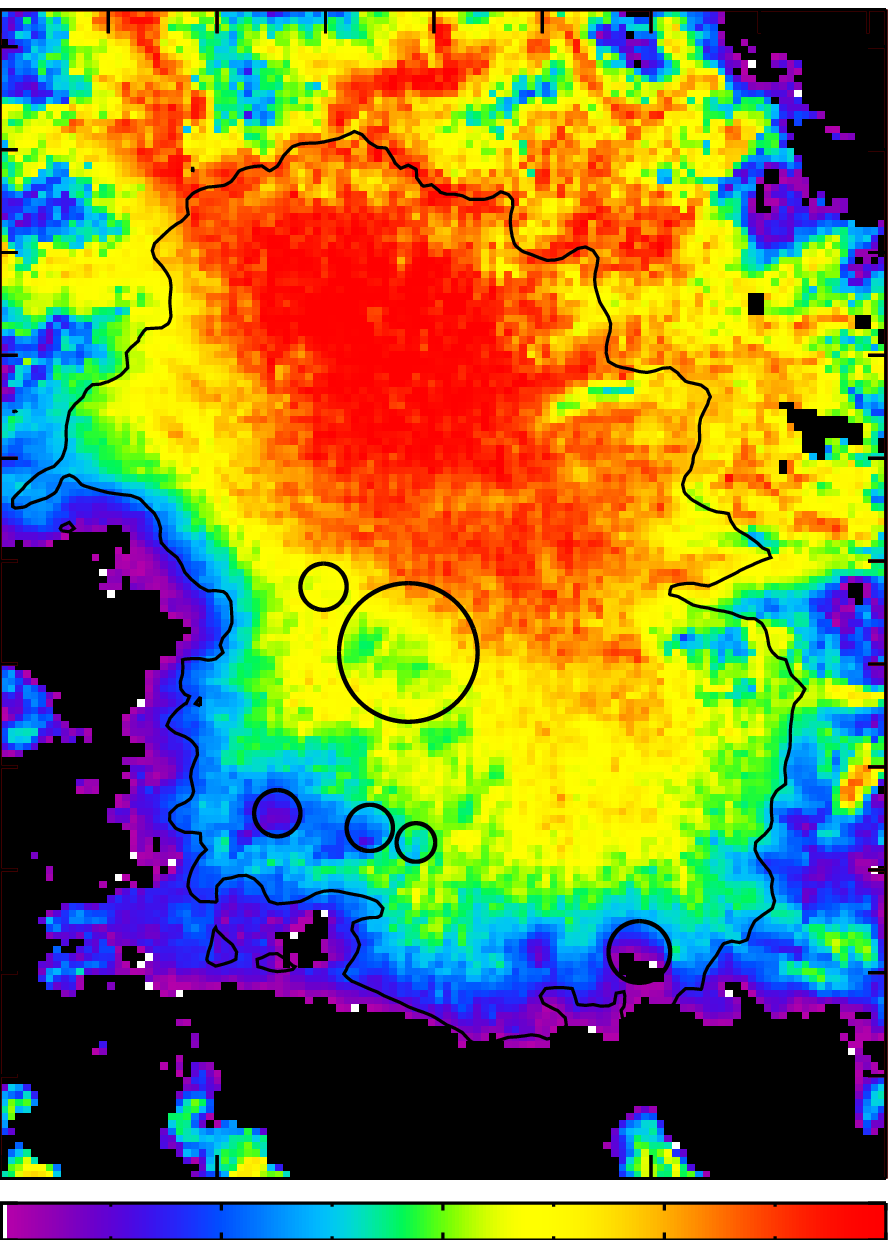}}
\vspace*{1.3cm}
\caption{Magnetic properties of UDs. In all panels we have 
marked the edges of the pore with contour lines and a few UDs with
circles. Tickmarks are separated by 1\arcsec. {\em Top:} continuum
image of the pore and its magnetoconvective structures (left); mean
linear polarization degree (middle); mean circular polarization
degree (right). {\em Bottom:} net circular polarization (left); 
magnetic field strength (middle); magnetic field inclination
(right). \label{fig5}}
\vspace*{.3cm}
\end{figure*}

Figure~\ref{fig5} shows maps of circular and linear polarization
signals, NCPs, field strengths, and field inclinations in the pore. 
The circles indicate selected UDs, and the dark lanes are named 
DL1 and DL2.

In linear polarization, the UDs stand out clearly above the umbral
background. The umbra has weak signals due to its nearly vertical
fields and the small heliocentric angle of the observations. UDs 5 and
7 get the higher LP values among the selected UDs. Interestingly, the
dark lanes DL1 and DL2 exhibit smaller degrees of linear polarization
than their corresponding UDs. In circular polarization, the UDs are
also the most prominent features, despite the fact that the umbral
background is now very intense. The dark lanes again show smaller
signals than their UDs. 

It is remarkable that many UDs have strong negative NCPs. The temporal
evolution of those signals is displayed in the lower panels of
Figure~\ref{fig3}. The NCP patches are shifted with respect to the
photometric center of the UDs, and tend to coincide with downflow
areas (e.g., UD4 and UD5) or with the border of an upflow patch (e.g.,
UD6). The dark lanes exhibit nearly zero NCPs, but the downflows at
their endpoints sometimes show large negative values (e.g., DL2). The
asymmetries of Stokes $V$ provide us with valuable information. As
discussed by \cite{solanki93}, the sign of the NCP is determined by 
the following rules:
\begin{eqnarray}
{\rm sign \, (NCP) } &=& - {\rm sign} \left( V_{\rm blue} \, \frac{ {\rm d}v_{\rm LOS}}{{\rm d}\tau} \, \frac{{\rm d}B}{{\rm d}\tau}
\right), \\
{\rm sign \, (NCP) } &=& - {\rm sign} \left( V_{\rm blue} \, 
\frac{{\rm d}v_{\rm LOS}}{{\rm d}\tau} \, \frac{{\rm d}|\cos \gamma|}{{\rm d}\tau}\right),
\end{eqnarray}
where $V_{\rm blue}$ represents the sign of the blue lobe of Stokes
$V$ (negative in our case), $B$ is the magnetic field strength
(positive by definition), and $v_{\rm LOS}$ is positive for downflows
and negative for upflows. We expect larger changes of $B$ with height
than of $\gamma$, hence the first relation above is likely to set the
NCP sign. The NCP patches coincide with downflows for the most
part. Since the bisector velocities increase toward the continuum,
${\rm d}v_{\rm LOS}/{\rm d}\tau > 0$ in those areas. Thus, to produce
a negative NCP it is necessary that ${\rm d}B/{\rm d}\tau$ be
negative, i.e., the magnetic field strength must {\rm decrease} as the
optical depth increases. This result lends support to the idea that
UDs are structures with significantly reduced fields near or just
below the continuum forming layers.

The interpretation of the observed polarization signals is difficult
because of the interplay of the various atmospheric parameters and the
existence of thermal effects. The field strength and inclination maps
of Figure~\ref{fig5}, however, show that many UDs are associated with
weaker and slightly more inclined fields than their surroundings. The
dark lanes have even weaker fields. The areas of weaker fields seem to
coincide in general with the locations of downflows but not with
upflows (the dark lanes are an exception, showing both upflows and
weaker fields). In the simulations, the downflows of the largest UDs
are sometimes associated with reversed magnetic polarities, but only
near $\tau = 1$ \citep{bhar10}. We do not observe opposite-polarity
fields in our observations.

Table~\ref{table3} quantifies the properties of the UDs marked in
Figure~\ref{fig5}. To allow comparisons, we also give the properties
of nearby points in the umbral background (UB). The magnetic field
strength of the UDs is considerably weaker than that of the umbra,
which peaks at about 2400~G. Differences are smaller when the comparison is
made with the immediate UB. In some cases, like UD4, the magnetic
field is as weak as 800 G. The dark lanes are the regions with weaker
fields within the UDs: DL1 and DL2 show strengths of 1000--1200~G,
while the bright parts of the corresponding UDs have fields of some
1500--1600 G. In the umbra, the field is inclined to the LOS by up to
160$^\circ$, as expected for vertical fields at an heliocentric angle
of $\sim$$30^\circ$. By contrast, the UDs exhibit inclinations closer
to $90^\circ$ and thus possess more horizontal fields. UDs 4 and 7 in
particular have the largest inclinations ($124^\circ$ and $121^\circ$,
respectively), perhaps because of their location near the edge of the
pore. Compared with the adjacent background, UDs show inclination 
differences of $3^\circ$--$17^\circ$.  The dark lanes have similar 
inclinations as their UDs, but higher than other unresolved UDs.

\begin{deluxetable}{lccccr}
\tablecolumns{6}
\tablewidth{8.5cm}
\tablecaption{{\rm Magnetic Properties of Selected UDs \label{table3}}}
\tablehead{
\colhead{Structure} & \colhead{B} & \colhead{$\gamma$} & \colhead{LP} & \colhead{CP} & \colhead{NCP}  \\
\colhead{} & \colhead{(G)} & \colhead{($\degr$)} & \colhead{(\%)} & \colhead{(\%)} & \colhead{(m\AA)}
}
\startdata
UD1 & 1610 & 139 & 4.0 & 9.8 &  $-1.5$ \\
DL1 & 1220 & 141 & 3.2 & 8.1 &  $-0.5$  \\
UB1 & 1900 & 150 & 2.1 & 8.2 &  $-0.4$ \\
\hline
UD2 & 1500 & 138 & 3.8 & 9.3 &  $-3$ \\
DL2 & 1020 & 141 & 2.8 & 7.4  & $0.8$  \\
UB2 & 1800 & 142 & 2.1 & 7.0 & $-0.8$ \\
\hline
UD4 & 800 &  124 & 4.1 & 5.0 & $-4$  \\
UB4 & 1200 & 134 & 2.5 & 5.0 & $-1.5$ \\
\hline 
UD5 & 1190 & 129 & 4.4 & 7.4 & $-9$  \\
UB5 & 1140 & 140 & 2.2 & 5.0 & $-2$  \\
\hline
UD6 & 1280 & 138 & 3.7 & 6.9 &  $-4$  \\
UB6 & 1450 & 141 & 2.2 & 5.8 & $-2.7$ \\
\hline
UD7 & 1070 & 121 & 4.9 & 6.0 & $-3$ \\
UB7 & 1480 & 140 & 2.3 & 5.6 & $-1.2$ 
\enddata
\tablecomments{
Column 1: number of UD and adjacent UB in Figure~\ref{fig5}. Column 2:
magnetic field strength. Column 3: inclination of magnetic field vector to
the LOS. Column 4: mean linear polarization degree.  Column 5: mean circular
polarization degree. Column 6: net circular polarization. \vspace*{.2cm}}
\end{deluxetable}

In summary, UDs exhibit weaker magnetic fields than the surrounding
umbra as well as more inclined field lines. These results confirm
previous works \citep[e.g.][]{hector04,rieth08a,sob09}. The new
information here is that of the dark lanes, whose magnetic properties
have not been determined earlier. We report that the dark lanes
exhibit an even weaker magnetic field than their associated UDs, 
up to 500 G less, as well as smaller LP, CP, and NCP signals. The field
inclination, however, does not differ from that of the UDs. Another
new result is the existence of enhanced NCP in the downflow patches
observed at the periphery of UDs. The sign of the NCP indicates that
the field strength decreases with depth in the photosphere.

\section{Discussion and conclusions}
\label{disc}

We have presented the first spectropolarimetric measurements of UDs 
at a resolution of 0\farcs14. Our observations reveal the existence of
substructures within UDs in the form of dark lanes. Only \citet{bhar07b}, 
\citet{rim08}, and \citet{sobpus09} have detected these structures 
before. \citet{rim08} estimated their size to be 0\farcs12 in G-band
images, right at the diffraction limit of the Dunn Solar Telescope,
while \citet{sobpus09} reported widths of less than 0\farcs14 from
broadband 602~nm filtergrams taken at the SST. Our observations also indicate sizes of 0\farcs14. At the
resolution of CRISP, however, not all UDs possess a dark lane.

In this paper, we have focused on the velocity field of UDs. One can
find in the literature a variety of works describing the morphological
properties of UDs, but very few attempts have been made to determine their velocities. However, very few attempts have been made to
determine their velocities. A good knowledge of the flow field of UDs
is important both to understand the energy transport in strongly
magnetized media and to discern between models of sunspot structure.

Measuring UD velocities is a difficult task, as recognized by several
authors \citep[e.g.,][]{rim04,hector04,simu06}. Our results, and those
from previous works, point to the existence of large LOS velocities
and reduced magnetic field strengths, but only in deep layers of the
photosphere --near the continuum forming region-- definitely deeper
than the layers traced by the core of many photospheric lines. For
example, \citet{rim04} finds upflows of more than 1 km s$^{-1}$ in the
\ion{C}{1} 538.0~nm line (whose core forms at approximately 40 km),
while he observes velocities of less than 300 ms$^{-1}$ in \ion{Fe}{1} 
557.6~nm (formation height of around 320~km). \citet{simu06} also 
mention the difficulty of observing flows and reduced field strengths 
in UDs because the surfaces of equal optical depth are locally elevated. 
Other factors such as the small sizes and the rapid evolution of 
these structures complicate the determination of their velocities 
and magnetic fields.

Observational studies have found upflows of 100 m\,s$^{-1}$
\citep{hector04}, 600 m\,s$^{-1}$ \citep{sob09}, 800 m\,s$^{-1}$ 
\citep{rieth08a}, and even 1000 m\,s$^{-1}$ \citep{rim04} in peripheral 
UDs.  Our data confirm the existence of relatively strong upflows of
around 1000 m\,s$^{-1}$, with peaks up to 1500 m\,s$^{-1}$. The upward
velocities observed in MHD simulations do not exceed 2000 m\,s$^{-1}$
at photospheric levels \citep{bhar10}.

The most relevant and novel result of this paper is the finding of
{\em downflows} at the edges of some UDs. If the UD has a dark lane,
then the downflows are observed at its endpoints. The downflow
velocities range from 400 m\,s$^{-1}$ to almost 1000 m\,s$^{-1}$. The existence of
downflows seems to fit the scenario proposed by \citet{simu06}, in
which UDs are the result of magnetoconvection in regions of reduced
field strengths with both upflows and return downflows. We believe
this is the first time that the predicted downflows are reliably
measured within UDs, mostly as a consequence of the high
resolution provided by CRISP and the SST. We have also examined the LOS velocities
at different intensity levels. Since the 80$\%$ intensity level shows
the strongest velocities and motions at other intensity levels are
gradually of lesser magnitude, a gradient of velocity with height
appears to exist. This suggests that the flows occur preferentially 
in the deep layers of the photosphere.

A comparison between our Figure~\ref{fig2}(f) and Figure 1(b) of
\citet{simu06} reveals some similarities, like the fact that the 
upflows are co-located with the centers of the UDs and their dark
lanes. Also in good agreement is the fact that the downflows are
observed around the UD structure, but most prominently at the
endpoints of the dark lanes. However, our measured velocities appear
to be slightly larger than those predicted by the simulations
\citep[see, e.g., Figure~16 of][]{bhar10}. In the same manner, the
temporal evolution shown in our Figure~\ref{fig3} may be compared with
Figure 5 of \citet{simu06}, where the highly dynamic and transient
nature of the UDs can be seen.

We have also determined the magnetic properties of UDs. New in this
work is the description of the properties of their substructure, the
dark lanes. UDs exhibit weaker magnetic fields and more inclined field
lines than the surrounding umbra, confirming previous analyses
\citep[e.g.][]{hector04,rieth08a,sob09}. We report a magnetic field
weakening of up to 500~G between UDs and their adjacent umbral
background, similar to the values given by \citet{rieth08a}.
According to our data, the magnetic field in the UDs is more
horizontal than in the immediate surroundings by as much as
$\sim$20$^{\circ}$. Before this work, the magnetic properties of the
dark lanes were largely unknown. We report that they exhibit an even
weaker field than their associated UDs, up to 500~G less. The dark
lanes also show smaller linear and circular polarization
signals. Their inclinations, however, do not differ from those of the
associated UDs. Another new result is the existence of enhanced NCP in
UDs. We observe tiny patches of NCP that tend to coincide with the
downflowing areas. In those regions, the NCP sign indicates a
reduction of the field strength towards deeper photospheric
layers. This again appears to confirm the results of numerical
simulations \citep{simu06}.

\acknowledgments

We are grateful to H\'ector Socas Navarro for valuable
discussions. Part of this work was carried out while one of us (A. O.) 
was a Visiting Scientist at the Instituto de Astrof\'{\i}sica de
Andaluc\'{\i}a. This research has been supported by the Spanish
Ministerio de Ciencia e Innovaci\'on through projects
ESP2006-13030-C06-02 and AYA2009-14105-C06-06, by Junta de
Andaluc\'{\i}a through project P07-TEP-2687, and by the Research
Council of Norway through grant 177336/V30. The Swedish 1 m Solar
Telescope is operated by the Institute for Solar Physics of the Royal
Swedish Academy of Sciences in the Spanish Observatorio del Roque de
los Muchachos of the Instituto de Astrof\'{\i}sica de Canarias. This
research has made use of NASA's Astrophysical Data System.

%{\em Facilities:} \facility{SST (CRISP)}

\end{document}